\documentclass[]{aastex631}

\newcommand{\af}[1]{\textcolor{black}{#1}} %

\graphicspath{{./}{figures/}}
\usepackage{soul}

\shorttitle{Constraints on the lunar deep interior from tidal deformation}
\shortauthors{Briaud et al.}

\begin{document}

\title{Constraints on the lunar core viscosity from tidal deformation}

\correspondingauthor{Arthur Briaud}
\email{briaud@geoazur.unice.fr}

\author{Arthur Briaud}
\affiliation{G\'eoazur, CNRS, Observatoire de la C\^ote d'Azur, Universit\'e C\^ote d'Azur, Valbonne, France}
\author{Agnès Fienga}
\affiliation{G\'eoazur, CNRS, Observatoire de la C\^ote d'Azur, Universit\'e C\^ote d'Azur, Valbonne, France}
\affiliation{IMCCE, Observatoire de Paris, PSL University, CNRS, Sorbonne Universit\'e, Paris, France}
\author{Daniele Melini}
\affiliation{Istituto Nazionale di Geofisica e Vulcanologia (INGV), Rome, Italy}
\author{Nicolas Rambaux}
\affiliation{IMCCE, Observatoire de Paris, PSL University, CNRS, Sorbonne Universit\'e, Paris, France}
\author{Anthony M\'emin}
\affiliation{G\'eoazur, CNRS, Observatoire de la C\^ote d'Azur, Universit\'e C\^ote d'Azur, Valbonne, France}
\author{Giorgio Spada}
\affiliation{Dipartimento di Fisica e Astronomia "Augusto Righi" {(DIFA)}, Alma Mater Studiorum, Università di Bologna, Bologna, Italy}
\author{Christelle Saliby}
\affiliation{G\'eoazur, CNRS, Observatoire de la C\^ote d'Azur, Universit\'e C\^ote d'Azur, Valbonne, France}
\author{Hauke Hussmann}
\affiliation{Deutsches Zentrum für Luft- und Raumfahrt (DLR), Berlin, Germany}
\author{Alexander Stark}
\affiliation{Deutsches Zentrum für Luft- und Raumfahrt (DLR), Berlin, Germany}
\author{{Vishnu Viswanathan}}
\affiliation{{Center for Space Sciences and Technology, University of Maryland Baltimore County, 1000 Hilltop Circle, Baltimore, MD 21250,USA}}
\affiliation{{NASA Goddard Space Flight Center, 8800 Greenbelt Road, Greenbelt, MD 20771, USA}}
\author{{Daniel Baguet}}
\affiliation{IMCCE, Observatoire de Paris, PSL University, CNRS, Sorbonne Universit\'e, Paris, France}

\begin{abstract}

We use the tidal deformations of the Moon induced by the Earth and the Sun as a tool for studying the inner structure of our satellite. Based on measurements of the degree-two tidal Love numbers $k_2$ and $h_2$ and dissipation coefficients from the GRAIL mission, Lunar Laser Ranging and Laser Altimetry on board of the LRO spacecraft, we perform Monte Carlo samplings for 120,000 possible combinations of thicknesses and viscosities for two classes of the lunar models. The first one includes a uniform core, a low viscosity zone (LVZ) at the core-mantle boundary, a mantle and a crust. The second one has an additional inner core. All models are consistent with the lunar total mass as well as its moment of inertia. By comparing predicted and observed  parameters for the tidal deformations we find that the existence of an inner core cannot be ruled out. Furthermore, by deducing temperature profiles for the LVZ and an Earth-like mantle, we obtain stringent constraints on the radius (500 $\pm$ 1) km, viscosity, $(4.5 \pm 0.8) \times10^{16}$ Pa$\cdot$s and the density (3400 $\pm$ 10) kg/m$^3$ of the LVZ. We also infer the first estimation for the outer core viscosity, (2.07 ± 1.03) × 10$^{17}$ Pa·s, for two different possible structures: a Moon with a 70 km thick outer core and large inner core (290 km radius with a density of 6000 kg/m$^{3}$), and a Moon with a thicker outer core (169 km thick) but a denser and smaller inner core (219 km radius for 8000 kg/m$^{3}$).

\end{abstract}


\section{Introduction} \label{sec:intro}

The Moon is the most well-known extraterrestrial planetary body thanks to observations from ground-based and space-borne instruments as well as lunar surface missions (see \citet{lognonne2003new,williams2009lunar,mazarico2010glgm,wieczorek2013crust,viswanathan2017inpop17a, viswanathan2019observational}). Data from Lunar Laser Ranging (LLR), magnetic, gravity, surface observations and seismic Apollo ground stations help us to quantify the  deformation undergone by the Moon due to body tides. These observations provide one of the most significant constraints that can be employed to unravel the deep interior (\citet{Williams2014b, williams2015tides}). Besides, gravity and LLR measurements provide good constraints on the moment of inertia as well as the total mass of the Moon (\citet{viswanathan2019observational}). The uncertainty
on the gravity field and the total mass, measurements have been significantly reduced by the Gravity Recovery and Interior Laboratory (GRAIL) mission.

The Moon deforms in response to tidal forcing exerted by, to first order, the Earth, the Sun and, by a lesser extent, by other planetary bodies. The forcing generates periodic variations of the degree-2 shape and gravity that depend on the internal composition and structure of the Moon. These changes in shape and gravity of the Moon are described by three geodetic parameters, called Tidal Love numbers (TLNs). 
The degree-two harmonic components of tidal deformation can be expressed by Love numbers {$k_{2}$} (potential perturbation), {$h_{2}$} (vertical displacement) and {$l_{2}$} (horizontal displacement). 
These low-degree TLNs are sensitive to the structure of the deep interior ($e.g.$ \citet{khan2004does}). Among them, the potential perturbation of TLN $k_{2}$ is related to the tidal changes of the moment of inertia and gravitational potential, and therefore is obtained from the precise measurement of the gravity field ($e.g.$ \citet{konopliv2001recent}) and rotation ($e.g.$ \citet{1994Sci...265..482D,2017NSTIM.108.....V}). 
Apart from {$k_{2}$}, the vertical displacement LN {$h{_2}$} has been estimated from LLR data (\citet{williams2015tides, viswanathan2019observational}), and independently, by the Laser Altimeter on board the Lunar Reconnaissance Orbiter (LRO) missions (\citet{2014GeoRL..41.2282M, 2021JGeod..95....4T}). These observations lead to a dichotomy of the TLN  $h_{2}$ of 0.04394$\pm0.0002$ for LLR and {$h{_2}$}=0.0386$ \pm 0.0022$ for analysis of Lunar Orbiter Laser Altimeter (LOLA) data (\citet{viswanathan2017inpop17a,2021JGeod..95....4T}).
These TLNs have uncertainties that have been significantly improved by the analysis of the GRAIL, LLR and LOLA data (\citet{Williams2014b,2014GeoRL..41.2282M,williams2015tides, viswanathan2019observational}). The horizontal displacement $l_{2}$ TLN will not be discussed here because it has not been estimated by any geodetic observation so far.

Apart from the geodetic constraints, the Moon and Mars ($e.g.$ \citet{zweifel2021seismic}) are the only other bodies besides the Earth for which seismic data are available. Seismic studies using the Apollo Passive Seismic Experiment (PSE) constrain the seismic wave velocity distribution and therefore give a glimpse of the lunar interior structure (\citet{2011PEPI..188...96G,weber2011seismic}). In principle, seismic data are the most informative data for deriving the density, rigidity and structure of any planetary interior. However, Moon-quakes are much weaker than earthquakes due to the lack of plate tectonics (\citet{shapiro2021bound,zhao2009deep}), and therefore they do not provide sufficient resolution on the deep interior of the Moon to detect all internal boundaries. The seismic wave velocities have been shown to be highly attenuated at a radius of $\approx$400km (P-waves) and $\approx$600km (S-waves) leaving the near-center structure uncertain (\citet{nakamura1983seismic, khan2000new, lognonne2003new}).  

Evidence from rotational dissipation (\citet{williams2001lunar}) and seismic velocity modeling (\citet{2011PEPI..188...96G, weber2011seismic}) suggest the presence of a fluid and dense core but do not reject the hypothesis of a differentiated core structure with a solid inner and an outer core. Other studies based upon geophysical constraints (\citet{khan2004does, matsumoto2015internal}) and the re-analysis of the Apollo seismic data suggested the existence of an attenuated region called the low-viscosity zone (LVZ) originating from a melting layer at the core-mantle boundary (\citet{khan2001new, weber2011seismic, harada2014strong,Rambaux14c}). This layer has a reduction in viscosity that can satisfy the seismic profiles, tidal parameters and dissipation coefficient ($e.g.$ \citet{weber2011seismic}). 
Several hypotheses exist about the present status of such a partially molten layer in the lunar mantle, inferring  different levels of hydration in the lowest part of the mantle (\citet{nimmo2012dissipation}). Such hydration if demonstrated will be crucial for a better understanding of mantle evolution and its exchange with the crust. The existence of a Moon inner core cannot be completely justified with only the Apollo seismic records. Even if most of the evolution scenarios are in favor of a differentiated core, the disappearance of the lunar magnetic field a few hundred thousand years after its formation addresses the question of the past lunar dynamo ($e.g.$ \citet{le2011impact}). The presence of an inner core will favor a dynamo mechanism while a pure fluid core will favor a strong convecting scenario (\citet{mighani2020end}).

In this paper, we aim at establishing new constraints on the lunar internal structure by generating a random ensemble of models and testing their compatibility with a range of observational constraints. In Sect. \ref{sec:NumericalApproach} we describe the semi-analytical approach used to estimate TLNs for a given scenario of interior structure. Models that are compatible with geodetic observations are identified and classified in homogeneous categories according to the procedure described in Sect. \ref{sec:methods}. The result of this statistical selection is presented in Sect. \ref{sec:prem_results}, while in Sect. \ref{sec:part5} we test the compatibility of our models with plausible hypotheses about the internal temperature profile of the Moon. In Sect. \ref{sec:discussion} we discuss the insights emerging from our analysis, before drawing our conclusions in Sect. \ref{sec:conclusions}.

\newpage

\section{Numerical approach and model setup}
    \label{sec:NumericalApproach}
We investigate the visco-elastic tidal deformation of the Moon by generating an ensemble of models and comparing their predicted tidal response with the most recent observational constraints (\citet{williams2005lunar,williams2015tides,matsumoto2015internal}). We use a modified version of the $ALMA$ code (\citet{spada2006using,spada2008alma}) to numerically estimate the TLN for a periodic forcing as a function of the assumed interior structure. As a reference model, we use the 1-D density and rigidity profiles of the Moon provided by \citet{Williams2014b}. However, we have made several changes ({see Sect.} \ref{sec:inputs}), especially for the lunar core following the assumptions of \citet{2011PEPI..188...96G} and \citet{weber2011seismic}. Our set of models is divided into two major categories: (1) a 4-layer structure assuming a uniform core, LVZ, mantle and crust, which shall refer to as $Category\,\it{4}$ and (2) a 5-layer structure where the core is further subdivided into a solid inner core and a viscous outer core, which will be called $Category\,\it{5}$. Both sets of models are set up with the same crust and mantle characteristics and include an LVZ at the base of the Moon mantle, as suggested in \citet{harada2014strong}. 

\subsection{$ALMA$ code}
    \label{sec:ALMAcode}

The LNs describe how a planetary body (in our case the Moon) deforms in response to a surface load or an external potential (in the present case, tidal forces) and how equipotential surfaces are consequently modified (\citet{love1909yielding,spada2008alma}). We use a semi-analytical code originally developed for studying Earth deformations, $ALMA$. The method behind $ALMA$ is explained in detail by \citet{spada2006using}, and \citet{2022GeoJI.231.1502M} introduced many new features to the up-to-date version of $ALMA$, $ALMA^3$ \footnote{available at \url{https://github.com/danielemelini/ALMA3}}. 
This version incorporates the tidal excitation and the possibility of defining a planetary profile with an elastic core. Here, we recall the most important characteristics and equations and briefly discuss how it {was} adapted to the case of periodic forcing \cite{2022GeoJI.231.1502M}.
$ALMA$ computes the Loading and Tidal Love Numbers (hereafter, LLNs and TLNs) for an incompressible, self-gravitating, radially layered planetary model. The approximation of incompressibility as assumed in the $ALMA$ code does not considerably affect our results due to the small size of the Moon. 
Moreover, \citet{kamata2012new} have obtained models showing the differences between the compressibility and incompressibility assumption on the LLNs $k’_2$ and $h’_2$. For periods shorter than 5 kyr, the incompressibility assumption does not critically affect the results.
In Appendix \ref{appendixA} are presented the comparisons of the Moon estimated TLNs with and without compressibility. In particular, Fig. \ref{fig:A2} and Table \ref{tab:TableA1} show the differences for the TLN $k_2$ and the quality factor. $ALMA$ {uses a multi-layered 1-D rheological profile as input} (i.e., radius, density, rigidity and viscosity). The original version of $ALMA$ is aimed at evaluating time-dependent LNs for a forcing term following a Heaviside time history. Within the framework of Viscoelastic Normal{ }Modes (VNMs), this is accomplished by computing the LNs in the Laplace domain and performing a numerical inverse Laplace transform in order to retrieve the LNs in the time domain.
$ALMA$ takes advantage of a non-conventional technique of Laplace inversion, the so-called "Post-Widder method" (\citet{post1930generalized,widder1934inversion}), introduced and benchmarked in \citet{spada2006using}, which allows to overcome most of the intrinsic limitations of VNMs. 
Since the Post-Widder method requires a numerical sampling of the LNs in the Laplace domain, $ALMA$ computes the Laplace-transformed solution of the equilibrium equations as follows: 
\begin{equation}
    \label{eq:ALMA1}
        \vec{x}(\added{R} \deleted{a},s)=f(s)[P_x W(s)J][P_b W(s)J]^{-1} \vec{b}
\end{equation}
with  
\begin{equation} \vec{x}(\added{R} \deleted{a},s)=(u,v,\phi)^t \label{eq:ALMA2}\end{equation}
where $u$, $v$ and $\phi$ are the vertical and horizontal components of the displacement and the incremental potential, respectively. In Eq. (\ref{eq:ALMA1}), $R$ is the planet radius, $s$ is the Laplace variable, $f(s)$ is the Laplace transform of the time-history of the forcing term, $W(s)$ is the (6$\times$6) matrix that propagates the solution from the core radius to the external surface, $P_x$ and $P_b$ are 3$\times$6 projection operators, $J$ is a $6\times 3$ matrix which accounts for the core-interface boundary conditions, and $\vec{b}$ is a vector expressing the loading or tidal boundary conditions at the surface ($e.g.$ \citet{sabadini1982polar,spada2008alma}).
The propagator $W$ has the form:
\begin{equation}
    \label{eq:ALMA3}
        W(s)=\prod_{j=L+1}^{1} Y_i(r_{j+1},s)Y_{j}^{-1}(r_{j},s)
\end{equation}
where the product index $j$ decreases from $j=L+1$ to $j=1$, $r_j$ $(j=1,...,L+2)$ is the radius of each interface, $L$ is the number of layers surrounding the central sphere, $r_1$ is the radius of the spherical layer, $r_{L+1}$ is the lithosphere-mantle boundary and $r_{L+2}=R$. For models assuming a uniform core ($Category\,4$), $r{_1}$ corresponds to the core-mantle boundary, while for models with a layered core ($Category\,5$), the core-mantle boundary is at $r_2$ and $r_1$ is the interface between the inner and outer core.


In Eq. (\ref{eq:ALMA3}), $Y(r,s)$ is the $6 \times 6$ fundamental matrix of the system of differential equations describing the radial part of the equilibrium and {Laplace} equation (\citet{spada2006using})  whose analytical form is given in \citet{sabadini1982polar}, while the elements of its inverse $Y^{-1}(r,s)$ are given by \citet{vermeersen1996analytical}. 
For an incompressible planet, the mantle rheology enters in $Y(r,s)$ through the {$s$}-dependent complex modulus (or effective shear modulus), whose form depends upon the kind of (linear) {rheological laws} assumed for the mantle (\citet{spada2008alma}). $ALMA$ can deal with several linear {rheological laws}; those used for our study are listed in Table \ref{tab:Table2}.

If the external forcing has a periodic time dependence, the solution can be obtained by setting  $f(s)=1$ and $s=i\omega$ in Eq.  (\ref{eq:ALMA1}), where $\omega$ is the forcing frequency and $i$ is the imaginary unit. In this case, the solution vector can be written as:
\begin{equation}
    \label{eq:ALMA4}
        \vec{x}(R,\omega)=[P_x W(i\omega)J][P_b W(i\omega)J]^{-1} \vec{b}
\end{equation}
The TLNs can then be obtained from the solution vector $\vec{x}(a,\omega)=(u(\omega),v(\omega),\phi(\omega))^t$ with the relation:
\begin{equation}\label{eq:lndef1}
\left(\begin{array}{c}
     u(\omega) \\ v(\omega) \\ \phi(\omega) 
\end{array}\right) =
\phi_{ext}
\left(\begin{array}{c}
     h/\gamma \\ l/\gamma \\ -(1+k)  
\end{array}\right) 
\end{equation}
where the {$\omega$}-dependence on the right-hand side has been left implicit, $\phi_{ext}$ is the potential of the tide-raising body and $\gamma$ is the surface gravity acceleration. According to ($e.g.$ \citet{wu1982viscous}), Eq. (\ref{eq:lndef1}) can be equivalently written as:
\begin{equation}\label{eq:lndef2}
\left(\begin{array}{c}
     h(\omega) \\ l(\omega) \\ k(\omega)
\end{array}\right) =
\left(\begin{array}{c}
     \xi u \\ \xi v \\ -1-\frac{\xi}{\gamma}\phi  
\end{array}\right) 
\end{equation}
where $\xi=m_m/a$ is the ratio between the Moon mass $m_m$ and its radius $a$.
Once the parameters of the layers have been set as inputs (i.e., radius, density, rigidity and viscosity of each layer), \af{$ALMA^3$} directly computes the real and imaginary parts of the LNs using Eqs. (\ref{eq:ALMA4}) and (\ref{eq:lndef2}).
{For a \af{Heaviside} forcing,} the short and long-term  asymptotic behaviours of the LNs correspond to the limits for $s\to\infty$ and $s\to0$ of Eq. (\ref{eq:ALMA1}), respectively, and are commonly referred to as "elastic" and "fluid" 
Love numbers ($e.g.$ \citet{hide1991earth}.
With $ALMA$ we are then able to estimate the tidal response of the Moon through frequency-dependent complex-valued TLNs. We also compute quality factors ($Q$), the corresponding dissipation coefficients, which are sensitive to the viscosity at the CMB interface.
To obtain $Q$, we first get from $ALMA$ the real and imaginary part of TLNs $k_2$ ($\Re$(k) and $\Im$(k)), respectively. The complex LN $k$ obtained from Eq. (\ref{eq:lndef2}) can be expressed as: 
\begin{equation}\label{eq:complexk}
    k= \Re(k) + i\Im(k)
\end{equation}
{which gives}
\begin{equation}
    |k|= \sqrt{[\Re(k)]^2 + [\Im(k)]^2}
\end{equation}
{so that} the tidal dissipation coefficient is calculated as follows,
\begin{equation}
    Q=\frac{|k|}{[\Im(k)]}.
\end{equation}
As described in \citet{williams2015tides} and in Sect. \ref{sec:obs}, the major periods of interest for the Earth-Moon system are $F$ =27.212 days and $\ell'$ = 365.260 days. Other periods were also discussed in previous studies but, as explained in \cite{williams2015tides}, the dissipation terms at 3-year and 6-year are more complicated to estimate from the LLR analysis and therefore most affected by uncertainties. 
Improving the 3-year and 6-year dissipating terms would lead to better constraints on the results. This is the reason why only the dissipation at 27.212 days and 365.260 days are accounted for in this study.  We thus compare the $ALMA$ outputs to the TLNs and dissipation coefficient obtained by \citet{williams2005lunar,williams2015tides,viswanathan2017inpop17a,2021JGeod..95....4T}.

\subsection{Observational constraints of the Moon}
    \label{sec:obs}

To delineate the frequency-dependence of the tidal parameters, we employed selenodetic observations of the mean radius ({$R$}), the mass ($M$), the normalized moment of inertia ({$C$}/$MR^{2}$), the TLNs of degree 2, $k_{2}$ and $h_{2}$, and the dissipation coefficient ($Q$) as reported in previous studies ($e.g.$ \citet{williams2005lunar, goossens2008lunar, matsumoto2015internal, williams2015tides}). The total mass, as well as the MoI, are derived from the GRAIL degree 2 gravity coefficient determination and LLR determination (\citet{Williams2014b}), at the Delaunay arguments $F$ of 27.212 days and $\ell'$ of 365.260 days (see Table \ref{tab:Table1}). We will use the estimations of dissipation coefficient and TLNs for these two frequencies as they were deduced from LLR observations and Fourier analysis by \citet{Williams2014b}.
\\
The estimates for {$k_{2}$} that will be used as first constraints in this work (see Table \ref{tab:Table1}) are mostly based on LLR and GRAIL data with values in between 0.0227 - 0.0310 corresponding to values proposed by \citet{williams2005lunar, Williams2014b}. Recent studies have shown that the observed $k_{2}$ values can be reduced to $0.02416\pm0.00022$ by taking the ellipticity of the gravity field into account (\citet{williams2015tides}). In addition, the {$k_{2}$} TLNs can be reduced to $0.0227\pm0.0025$ taking into account the fluid core oblateness (\citet{1994Sci...265..482D, williams2001lunar, konopliv2006global, williams2005lunar, williams2006lunar,2017NSTIM.108.....V}) .

In addition, the TLN {$h_{2}$} is also considered. However, due to large discrepancies reported by \citet{2017NSTIM.108.....V} and \citet{2021JGeod..95....4T},  divergences between the Lunar Laser Altimeter (LOLA) and the LLR-derived values are observed. We use the admitted range of $h_{2}$ (Table \ref{tab:Table1}) for constraining our results. 

\begin{table}
\caption{Selenodetic data used to constrain the modeled interior of the Moon. The Delaunay arguments $F$ and $\ell'$ correspond to periods defined by \citet{williams2015tides} of 27.212 days and  365.260 days, respectively.  [1] \citet{williams2005lunar}; [2] \citet{Williams2014b}; [3] \citet{williams2015tides}; [4] \citet{2017NSTIM.108.....V}; [5] \cite{2021JGeod..95....4T}; [6] \cite{goossens2008lunar}; [7] \cite{matsumoto2015internal}.}

    \centering
        \begin{tabular}{l c c c c c}
\hline
\af{{Data}} & {Symbol} & Value & 3-$\sigma$ & Reference\\
\hline
\hline
        Mean radius (km) & $R$  & 1737.1 &  & [3]\\
        Total mass (kg) & $M$ &  7.34630 $\times$ $10^{22}$ & $\pm$ 0.00264 $\times$ $10^{22}$ &  [3] \\
        Moment of Inertia & $C/MR^{2}$  & 0.393112 & $\pm$ 3.6 $\times$ $10^{-5}$ & [2, 7] \\
        Potential perturbation & $k_{2}$  &  0.02346 & $\pm$ 2.2 $\times$ $10^{-3}$ & [1, 6] \\
        Vertical displacement & $h_{2}$  &  0.0386 - 0.0430 & -- & [4,5] \\ 
        Monthly dissipation & $Q_F$ & 38 & $\pm$ 12 & [2] \\
        Yearly dissipation & $Q_{\ell'}$ & 41 & $\pm$ 27 & [2] \\
        Monthly libration & $(k_2/Q)_{F}$ & 6.4 $\times$ $10^{-4}$ & $\pm$ 4.5 & [3] \\
        Yearly libration & $(k_2/Q)_{\ell'}$ & 6.2 $\times$ $10^{-4}$ & $\pm$ 4.2 & [3] \\
            \hline
        \end{tabular}
        \label{tab:Table1}
\end{table}

\subsection{Geophysical inputs}
    \label{sec:inputs}

\begin{figure}

        \centering
            \includegraphics[scale=0.6]{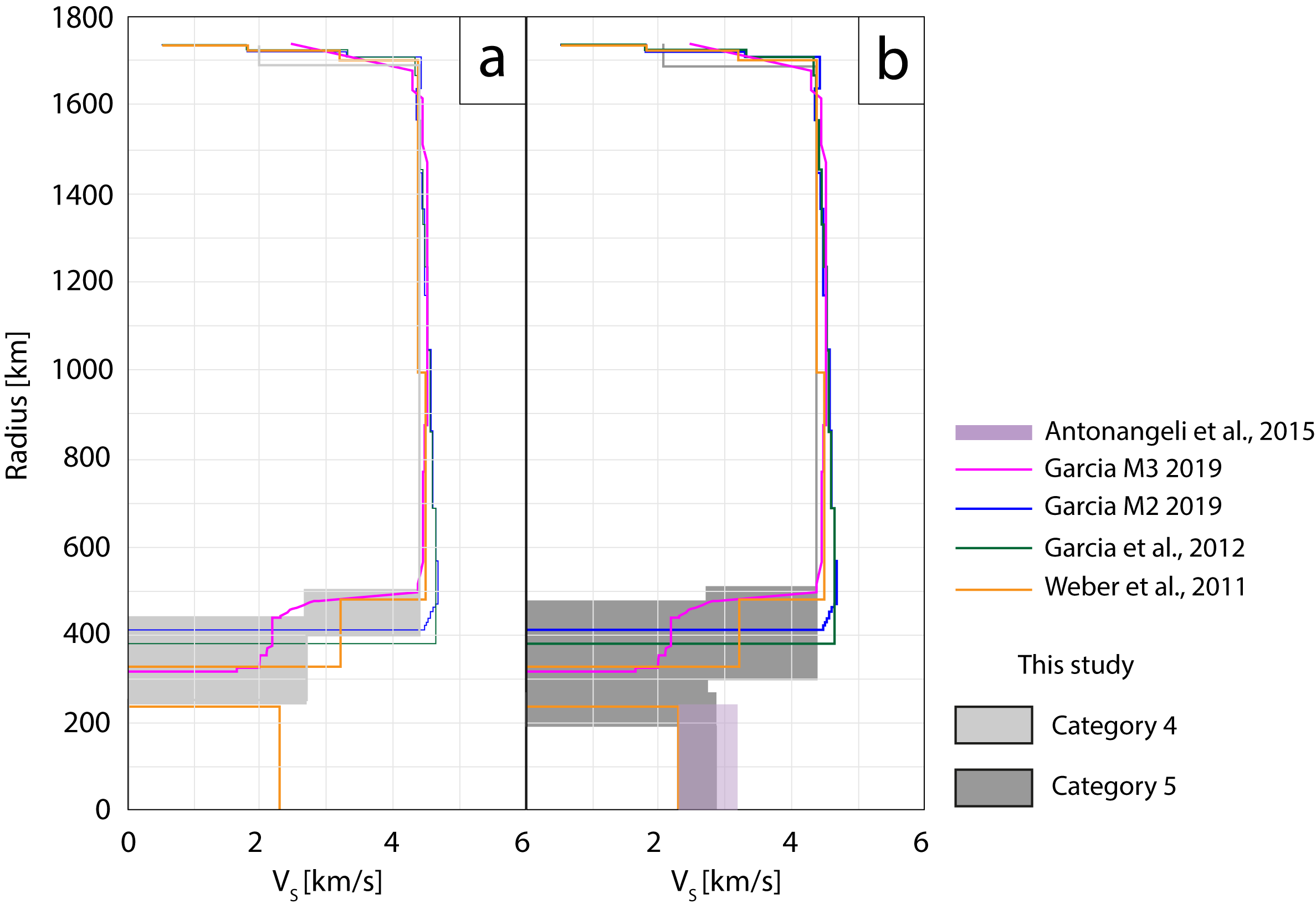}
    \caption{Range of 1-D V$_s$ profiles used to the $Category\,4$ (a) and $Category\,5$ (b). Colored lines represent models from previous studies of \citet{Antonangeli3916,2011PEPI..188...96G,garcia2019lunar,weber2011seismic}. }
\label{fig:1Dprofiles}
\end{figure}

As inputs for $ALMA$, we need to implement 1-D profiles that describe the hypothetical interior structure of the Moon. As mentioned in Sect. \ref{sec:ALMAcode}, the radius, the density, the rigidity and the viscosity are required to compute TLNs. Table \ref{tab:Table2} shows the structure of the 1-D profiles {that} we considered, the variability ranges of input parameters (density, rigidity, viscosity) and the rheological laws assumed in each layer. Our 1-D profiles (see Fig. \ref{fig:1Dprofiles}) are based on the seismologically{-}derived density and wave velocity (\citet{weber2011seismic, 2011PEPI..188...96G,garcia2019lunar}). The crust and the mantle are well-constrained thanks to the seismic studies from \citet{gagnepain2006seismic, weber2011seismic}{, see also} \citet{viswanathan2019observational,tan2021tidal}. 
So we assign to these two layers constant rheological parameters as listed in Table \ref{tab:Table2}. Previous studies (\citet{harada2014strong,harada2016deep}) have shown that the attenuation of the seismic waves in the deep interior is expected to be consistent with a viscosity reduction at the core-mantle boundary. This viscosity contrast may explain the frequency dependence of the dissipation coefficient (\citet{Rambaux14c,harada2014strong, matsumoto2015internal,harada2016deep}). For this LVZ layer, we thus consider a specific density and rigidity while we explore a wide range of viscosity values to investigate its effect on the TLNs and dissipation coefficient (see Table \ref{tab:Table2}). With our method, we expect to obtain a {strong} constraint of the LVZ viscosity as stated by \citet{harada2014strong}. The core remains the most uncertain part of the lunar interior. Hence, we consider four-layer models ($Category \, 4$) with only a Newtonian core and five-layer models ($Category \, 5$) including a solid inner core and an outer core according to \citet{weber2011seismic}, (see Fig. \ref{fig:1Dprofiles}).  
From \citet{weber2011seismic} and \citet{viswanathan2019observational} reference models, we then vary the parameters listed in Table \ref{tab:Table2} exploring the space of possible models compatible with the observational constraints given in Table \ref{tab:Table1}. We assign to each model a radial viscosity profile following the values listed in Table \ref{tab:Table2}. According to \citet{ross1986tidal}, the response of the Earth to lunar tides has been calculated by matching the dissipation coefficient in the mantle from free oscillations. 
This computation gave a viscosity of $10^{21}$ Pa$\cdot$s for the Earth mantle, which is in agreement with the mantle viscosity inferred from post-glacial rebound \cite{turcotte2002geodynamics}. We then adopt the same viscosity for the lunar mantle across all models. In contrast, we vary the LVZ and the outer core viscosity values assigning to each model a random value within the ranges listed in Table \ref{tab:Table2}. 
Compared to \citet{williams2001lunar,harada2014strong, harada2016deep} who have chosen an inviscid fluid core rheology, we decided to use a Newtonian rheology for taking into account the core viscosity. Indeed, previous studies of \citet{secco1995viscosity} have argued different ranges of viscosity of the Earth outer core. 
So, by analogy, we explore from very low-viscosity to quasi-elastic core behaviour for testing the TLNs and dissipation coefficient sensitivity to the core viscosity. The crust and the inner core of the Moon are assumed to be elastic. For the crust, this choice is supported by evidence of viscous relaxation of the crater topography (\citet{namiki2009farside}) reaching the elastic limit (that is a viscosity greater than $10^{27}$ Pa$\cdot$s). The {solid} inner core is considered to have a rigidity similar to the one in \citet{weber2011seismic}.

\begin{table}
        \centering
\caption{General 1-D profiles of the Moon interior. Values in brackets [ ] indicate the range of parameters that vary randomly and uniformly between models. [**] represents the values that depend on the random uniform distribution of the radius. Please note that the core layer is present only in $Category\,4$, while the outer core and inner core are included only in $Category\,5$ (see Sect.\ref{sec:NumericalApproach}).}
     \begin{tabular}{l c c c c c c}

\hline
    Layer & Radius & Density & Rigidity & Viscosity &  Rheology\\
    \hline
    \hline
    Unit & km & kg/m$^{3}$  & Pa & Pa$\cdot$s &  -- \\
    \hline
        Crust & 1737.1  & 2700 &  $1.60 \times 10^{10}$  & --  &  Elastic\\
        Mantle & 1690 & 3380 & $6.56 \times 10^{10}$  & $1 \times 10^{21}$ & Maxwell\\
        LVZ & [450 : 700] & [**]  & $2.48 \times 10^{10}$& [$1 \times 10^{1}$ : $1 \times 10^{30}$] & Maxwell\\
        Core & [250 : 450] & [**] & 0 & [$1 \times 10^{1}$ : $1 \times 10^{30}$] & Newton\\
        Outer core& [250 : 450] & [**] & 0 & [$1 \times 10^{1}$ : $1 \times 10^{30}$]  & Newton\\
        Inner core & [120 : 250] & [**] & $4.23 \times 10^{10}$ & --  & Elastic\\
\hline
            \end{tabular}

\label{tab:Table2}
\end{table}

\newpage
\section{Method}\label{sec:methods}

The approach for this study is to randomly vary the input parameters required to estimate the measures for the lunar deformation (frequency-dependent TLN and dissipation coefficient) and compare $ALMA$ outputs (as given in Sect. \ref{sec:ALMAcode}) to the observational constraints (see Sect. \ref{sec:obs}). First, we select models that satisfy the lunar total mass and MoI (see Table \ref{tab:Table1}). Second, we compute tidal deformation for the selected lunar models and compare predictions with the six observed constraints from Table \ref{tab:Table1}. We use the 3-$\sigma$ quantiles of the Weighted Residual Sum of Squares (WRSS) distribution as criteria for selecting models compatible with the observations (Table \ref{tab:Table1}).

\subsection{Step 1: Total mass and moment of inertia criteria}
\label{sec:MassMoI}
    
In order to explore a wide range of hypothetical models of the lunar interior, we use a uniform random distribution for the radius and viscosity (Table \ref{tab:Table2}). The distribution of the density is {varied} according to the random distribution of the radius of each layer. Hence, each varying density layer respects the mass conservation constraint.
For constraining random radii that follow a uniform distribution and the deduced densities, we use the observational constraints of the total mass of the Moon and its MoI. Hence, the models that are considered are only those that agree within the 3-$\sigma$ quantiles of the observed mass and MoI listed in Table \ref{tab:Table1}. 

For each simulation, we first calculate the total mass of the Moon according to:
\begin{equation}\label{eq:mass}
    M = \frac{4}{3} \pi \sum_{j=0}^{L+1} \rho_j 
    \left( r_{j+1}^3 - r_j^3 \right) 
\end{equation}
where $\rho_j$ is the density of the \af{$j^{th}$} layer ($j=1,\ldots,L+1$), $\rho_0$ is the density of the core or inner core and $r_0=0$. The MoI for each model is estimated as follows:
\begin{equation}
    \label{eq:MoI}
    C = \frac{8\pi}{15} 
    \sum_{j=0}^{L+1} \rho_j \left(
    r_{j+1}^5 - r_j^5 \right).
\end{equation}
Thus, the normalized MoI, $\widetilde{C}$, for each model is obtained from {Eqs.} (\ref{eq:mass}) and (\ref{eq:MoI}) as: 
\begin{equation}
    \label{eq:normMoI}
 \widetilde{C} = \frac{C}{MR^{2}} 
\end{equation}
where $R$ is the mean radius of the Moon, in our case, the crust radius (see Table \ref{tab:Table2}).
\\
After filtering out our simulation profiles according to the total mass and the MoI, we use $ALMA$ to compute TLNs subsequently employed to constrain the Moon inner structure from the $3$-$\sigma$ observational uncertainties (Table \ref{tab:Table1}). By doing so, we ensure that all selected modeling are consistent, not only, with the observational constraints in TLNs and dissipation coefficient (see Sect. \ref{sec:WRSS}), but also in mass, MoI and{ }Vs profiles.

\subsection{Step 2: WRSS filtering}
\label{sec:WRSS}

\subsubsection{Construction of the Weighted Residual Sum of Squares distribution}

\begin{figure}[ht]

        \centering
            \includegraphics[scale=0.6]{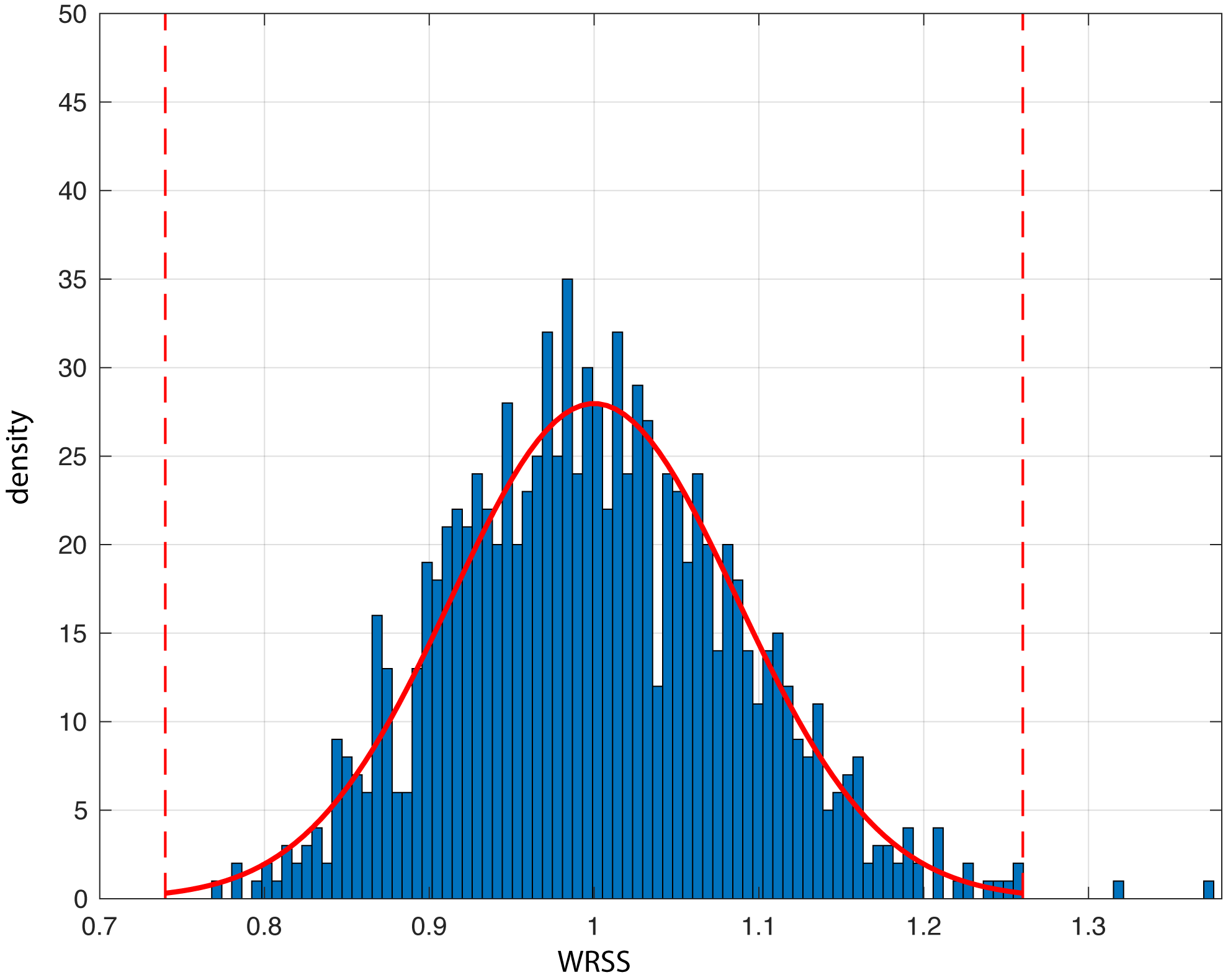}
                \caption{Modeled Weighted Residual Sum of Squares distribution (blue histogram). The red line corresponds to the Gaussian fit used to compute the standard deviation employed to decipher the simulated observable from the 3-$\sigma$ uncertainties (dashed red lines).}
\label{fig:chi2woBC}
\end{figure}

We study the instrumental noise variability of the constraints obtained from lunar observations described in Sect.\ref{sec:obs} and given in Table \ref{tab:Table1}.
To do so, we generate a set of simulated observables composed by the {$k_2$} {and} {$h_2$} TLNs and dissipation coefficient{s} at 27.212 and 365.260 days periods obtained by adding a Gaussian noise to the reference observables. The Gaussian noise standard deviation corresponds to the observable 3-$\sigma$ uncertainties as given in Table \ref{tab:Table1}.
We operate 1000 samplings and for each of them we compute the differences between the simulated observables and the reference ones as well as the Weighted Residual Sum of Squares (WRSS) for the 6 observables as follows:
\begin{equation}
    \label{eq:chi2}
        WRSS  = \frac{1}{N} \sum_{i=1,N} \left(\frac{(O-{\af{S}})_{i}}{\sigma_{i}}\right)^2
\end{equation}
where $(O-S)_{i}$ is the difference between the $i^{th}$ observable taken as reference (presented in Table \ref{tab:Table1}), $O$, and the simulated ones, $S$. $\sigma_{i}$ is the 3-$\sigma$ uncertainty as given in Table \ref{tab:Table1} and $N$ is the number of observables. We obtain an experimental WRSS distribution as presented in Fig. \ref{fig:chi2woBC}. From this empirical distribution, we can estimate the probability of a WRSS being explained by the instrumental uncertainties associated with the reference values.
We derive the quantiles corresponding to the 3-$\sigma$ of the WRSS  distribution after fitting a log-normal profile. We can then use a confidence interval $[$WRSS $_{min}$: WRSS $_{max}]$  that contains 99.7$\%$ of the distribution for selecting the lunar interior models compatible with the observations of the tidal deformation. The selected models are the ones for which the WRSS between modeled and observed tidal parameters belongs to the interval $[$WRSS $_{min}$: WRSS $_{max}]$ for the two periods of interest.

\subsubsection{Filtering}
    \label{sec:filtering}

\begin{figure} [ht]

        \centering
            \includegraphics[scale=1]{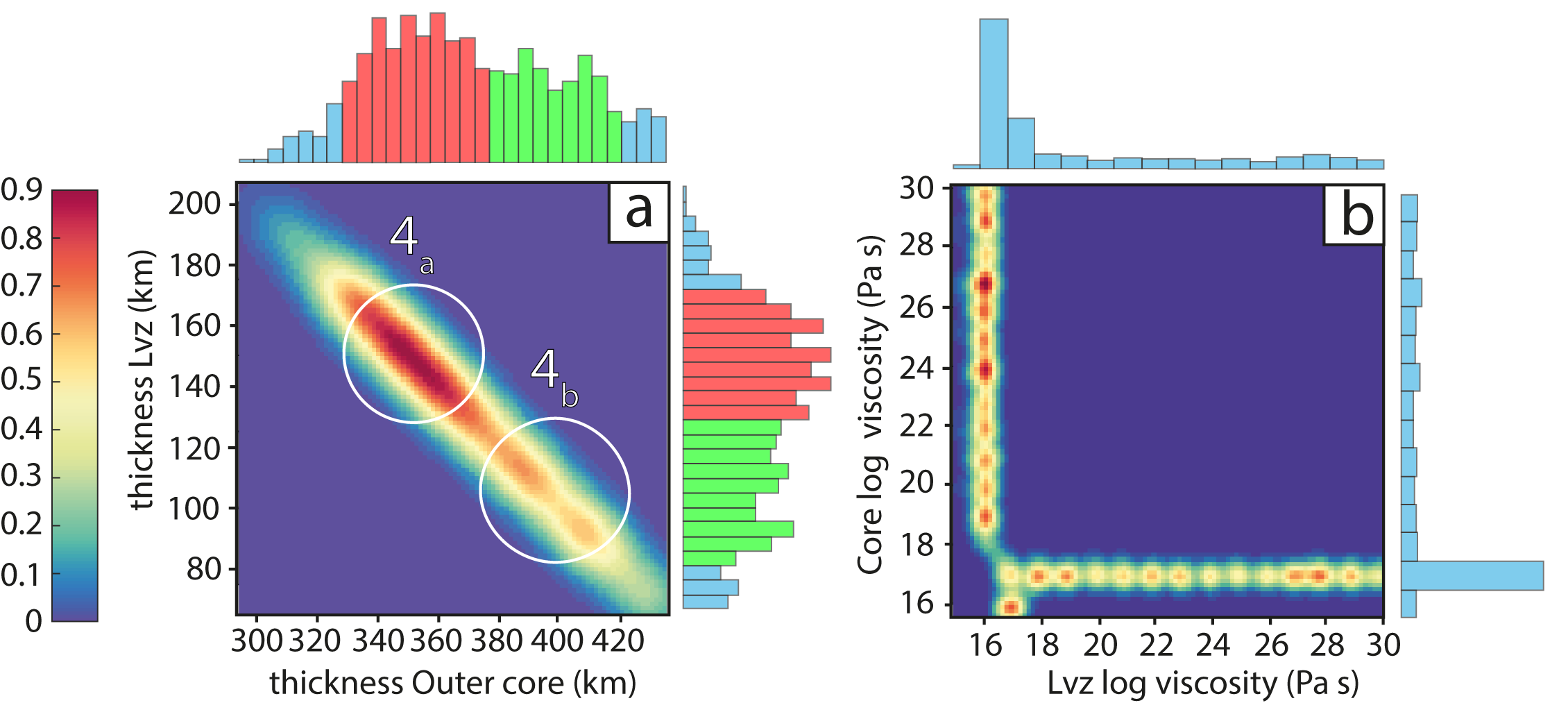}
                \caption{2-D probability distributions of the 4-layer model  (without inner core). On each panel, the color scale indicates the normalised probability distribution. White circles indicate the two sub-categories ($Category\,4_a$ and $4_b$), with their respective standard deviation, resulting from the $k-means$ algorithm. Histograms on panels a and b correspond to the distribution of the thicknesses and viscosities, respectively. The red and green histograms correspond to the selected models for $Category\,4_a$ and $Category\,4_b$, respectively.
        }
\label{fig:2Dhistoa}
\end{figure}

\begin{figure}

        \centering
            \includegraphics[scale=1]{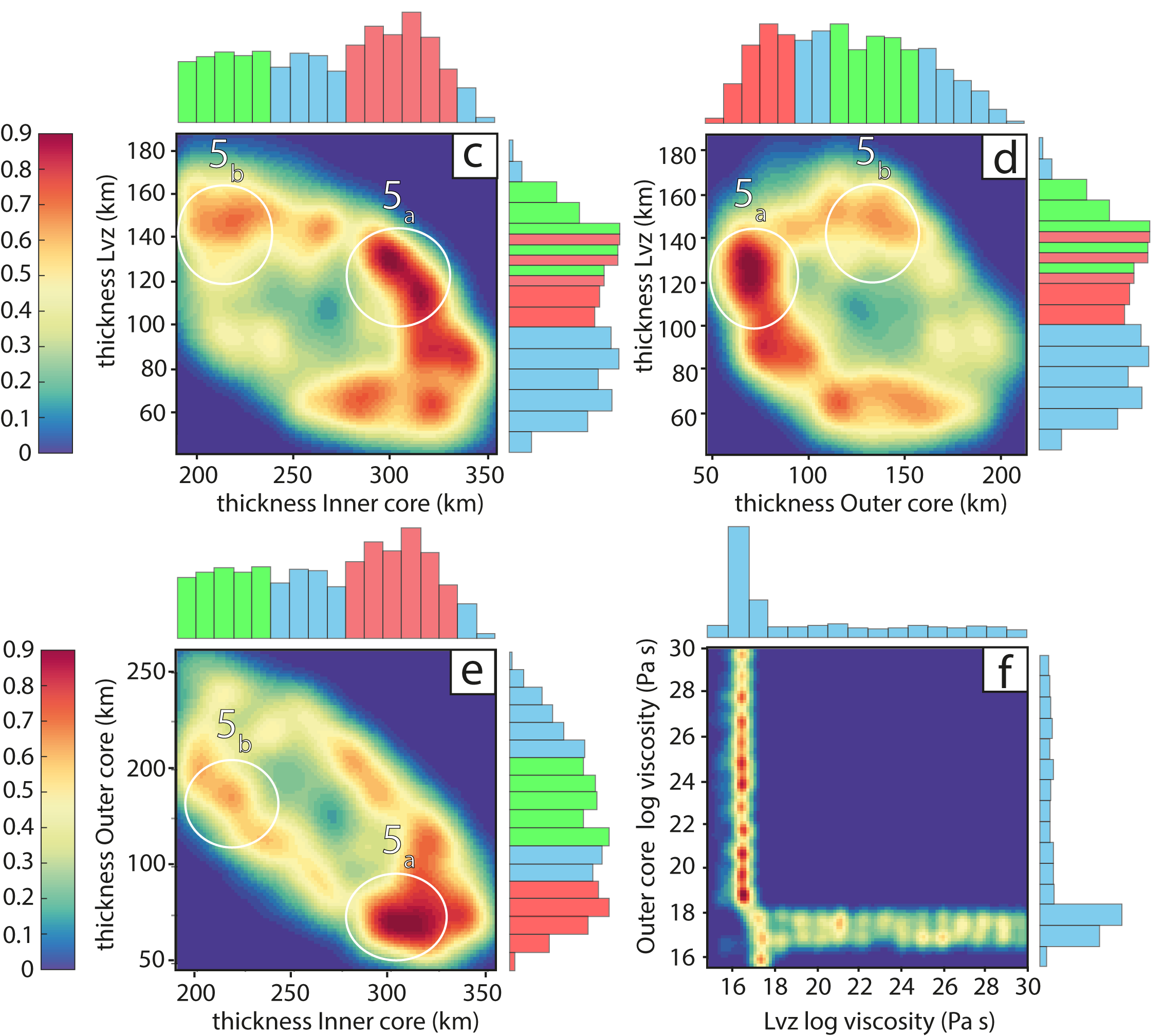}
                \caption{2-D probability distributions of the 5-layer model (with inner core). In each panel, the color scale indicates the normalised probability distribution. White circles  indicate the two sub-categories ($Category\,5_a$ and $\,5_b$), with their respective standard deviation, resulting from the $k-means$ algorithm. Histograms on panels c, d, e and f correspond to the distribution of the thicknesses and viscosities, respectively. The red and green histograms represent the selected models for $Category\,5_a$ and $Category\,5_b$, respectively.
        }
\label{fig:2Dhistob}
\end{figure}

As described previously, for each simulation, we compute the WRSS as defined in Eq. (\ref{eq:chi2}) and we keep then only models whose WRSS is encompassed in the interval $[$WRSS $_{min}$: WRSS $_{max}]$.
In the following, this test will be called WRSS filtering. We apply this filtering and extract the 2-D histograms for the selected models considering the distribution of layer thicknesses and viscosities. 
Figs. \ref{fig:2Dhistoa} and \ref{fig:2Dhistob} show the 2-D probability distributions for the 4-layer model and the 5-layer models, respectively.

\subsection{Step 3: K-means clustering algorithm}
    \label{sec:kmean}

As one can see in Fig. \ref{fig:2Dhistoa} and Fig. \ref{fig:2Dhistob}, concentrations of models exist in the parameter space for some ranges of thicknesses.
A fundamental issue that arises then from the 2-D probability distribution is the clustering problem $e.g.$, the pattern recognition of statistically significant model clusters in the parameter space.
In this study, the clustering problem is defined as the problem of finding homogeneous sub-categories of data points from a given category of models. Each of these sub-categories is called a cluster and is defined as a region in which the density of selected models is locally higher than in other regions of the 2-D probability histograms (Figs. \ref{fig:2Dhistoa}-\ref{fig:2Dhistob}). 
To identify the statistically significant clusters, we employed the $k$-means algorithm, whose details are illustrated in Appendix \ref{appendixB}.
For both $Category \,4$  and $Category \,5$, we identify the relationships between the thickness of the layers that vary the most between models as described in Table \ref{tab:Table2}. For the $Category \,4$ (without inner core) the layer thicknesses that vary are the ones of the LVZ and of the core (see Fig. \ref{fig:2Dhistoa}). In contrast, $Category \,5$ gets three layer thicknesses that vary. Hence, we built three 2-D marginal histograms to decipher relationships between layers, shown in Figs. \ref{fig:2Dhistob}-c,d,e.
Based on the Silhouette parameter estimation described in Appendix \ref{appendixB}, we find for $Category \,4$  two statistically significant clusters corresponding to sub-categories $4_a$ and $4_b$ in Table \ref{tab:Table3}. Fig. \ref{fig:2Dhistob} displays a more complex distribution of thicknesses between layers (see also Table \ref{tab:Table2}) for the $Category \,5$ set of models. Nevertheless, also in this case we find two statistically significant sub-categories $5_a$ and $5_b$, presented in Table \ref{tab:Table3}. The Silhouette parameters (Appendix \ref{appendixB}, Figs. \ref{fig:ann1}-\ref{fig:ann3}) do not identify statistically significant clusters for the viscosity probability distribution of the two $Categories$ 4 and 5, respectively (Figs. \ref{fig:2Dhistoa}-b and \ref{fig:2Dhistob}-f).

\begin{table}
        \centering

            \caption{Layer thickness for sub-categories identified by the $k-means$ clustering algorithm. Uncertainties correspond to the $3$-$\sigma$ standard deviation.
            The superscripts $^1$ and $^2$ refer to the 4-layer and 5-layer modeling, respectively.}
            \begin{tabular}{c c c c c c}
\hline
    Category & & & Thicknesses & \\
   & LVZ & Core$^1$ & Outer core$^2$ & Inner core$^2$ \\
    \hline
   & km & km & km & km \\
\hline
\hline
     4$_a$ &  154 $\pm$ 24 & 355 $\pm$ 25 & -- & --\\
     4$_b$ &  101 $\pm$ 23 & 386 $\pm$ 23 & -- & --\\
     5$_a$ &   123 $\pm$ 22 & -- & 76 $\pm$ 14 & 304 $\pm$ 26\\
     5$_b$ &   141 $\pm$ 27 & -- & 142 $\pm$ 28 & 226 $\pm$ 23\\
     5$_c$ &   127 $\pm$ 33 & -- & 102 $\pm$ 29 & 277 $\pm$ 46\\
\hline
            \end{tabular}
                \label{tab:Table3}

\end{table}

\newpage

\section{Results}
\label{sec:prem_results}

We performed about 120,000 simulations to determine the lunar interior structure using TLNs and dissipation coefficient constrained by LLR and LOLA observations, the total mass, the MoI and the V$_s$ profiles. In this section, we present the results of our statistical analysis (see Sect. \ref{sec:WRSS} - \ref{sec:kmean}) for the two categories of models defined in Sect. \ref{sec:inputs} (Tables \ref{tab:Table2} and \ref{tab:Table3}). 

In what follows, to quantify the distribution of model parameters we use the 25$^{th}$, 50$^{th}$ and the 75$^{th}$ percentiles. The 50$^{th}$ percentile, also known as the median, splits the data set in two equal parts meaning that half of the models lead to values below the median value and a half lead to values above the median. The 75$^{th}$ percentile identifies the value at which 75$\%$ (25$\%$) of the models lead to lower (higher) values. We will present our results using the notation $a^{b}_{c}$, where $a$ is the median ($50^{th}$ percentile) while $b$ and $c$ are the $75^{th}$ and $25^{th}$ percentiles, respectively.

\subsection{Sensitivity analysis of the TLNs and dissipation}

\begin{figure}[h]
        \centering
            \includegraphics[scale=0.8]{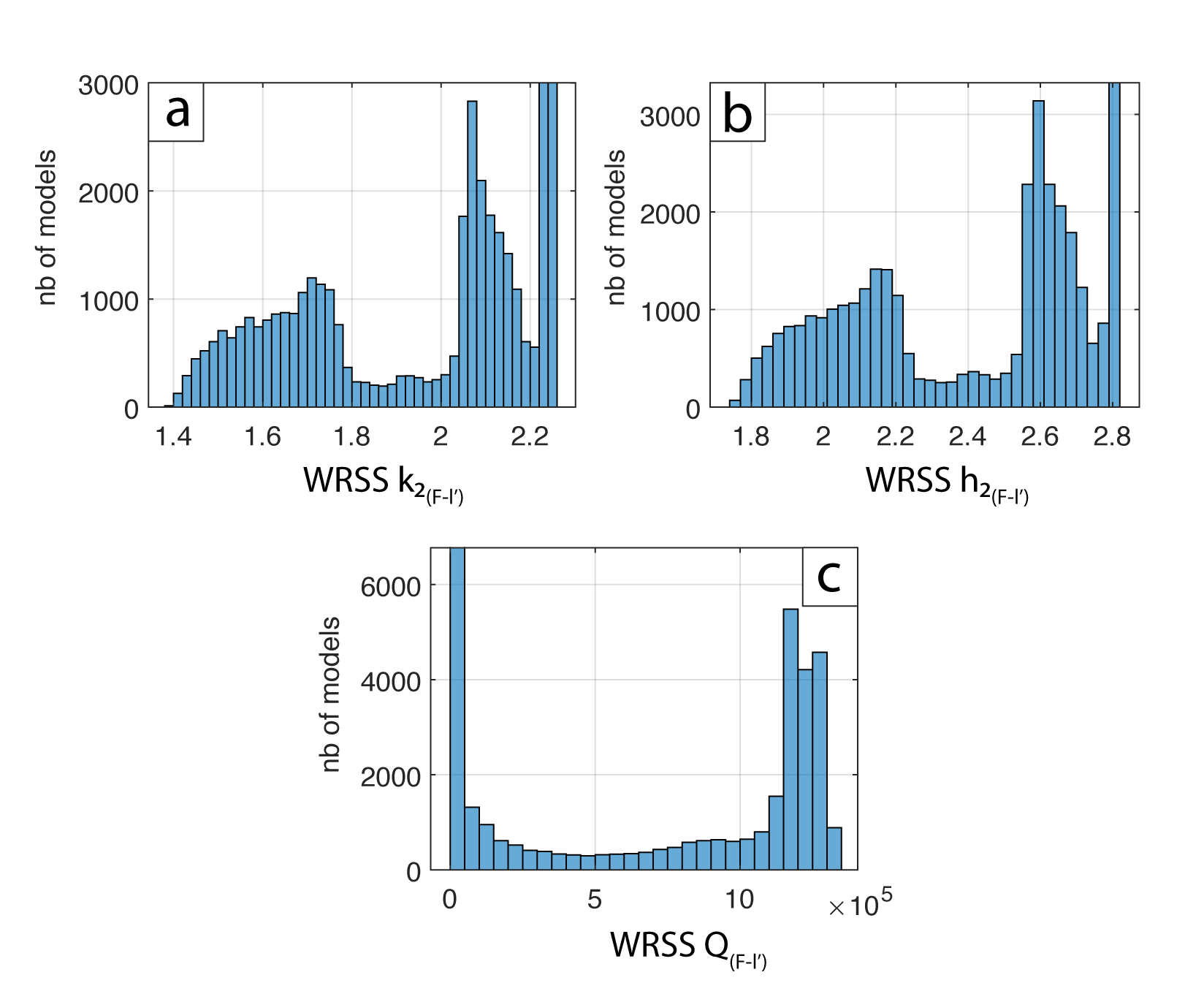}
                \caption{1-D histograms of WRSS for TLNs $k_2$ (a), $h_2$ (b) and quality factor $Q$ (c). For each parameter, WRSS is computed by considering both periods of interest, 27.212 days and 365.260 days, respectively.}
\label{fig:Filtering}
\end{figure}

We apply the WRSS filtering considering the two periods mentioned in Sect. \ref{sec:NumericalApproach}. Fig. \ref{fig:Filtering} shows the distribution of WRSS for the selected models, showing separately the contributions from TLNs $k_2$ (a) and $h_2$ (b) and from dissipation coefficient $Q$ (c). The WRSS range lies between 1.4 and 2.2 for $k_2$, between 1.8 and 2.8 for $h_2$ and spans over 5 orders of magnitude for $Q$. The differences between the intervals of variations for the TLNs and $Q$ suggest that dissipation is the parameter that controls the selection of the models. Applying the WRSS filtering extracts 962 (1.60\%) and 1126 (2.20\%) models respectively for the $Category\,4$ and $Category\,5$ sets, each of them consisting of 60000 models.

\subsection{4-layer ($Category\,4$)}
    \label{sec:Category1}

The models of $Category\,4$ have four layers including a crust, a mantle, an LVZ and a core. Over the 60,000 models, the WRSS filtering (Sect. \ref{sec:WRSS}) extracts 962 models (1.60\%) that match the observational constraints (Sect. \ref{sec:obs}).
We also consider the impact of using an Andrade model for the mantle instead of a Maxwell model. After filtering both Maxwell and Andrade models, we obtain 1178 models for the Andrade model against 962 models for the Maxwell model. The difference over the total number of models is thus about 0.36$\%$ after the WRSS filtering. The impact of considering Andrade instead of Maxwell for the mantle rheology is then negligible in this study. Other rheological tests can be found in Appendix \ref{appendixC}.
The resulting 1-D profiles are given in Table \ref{tab:Table4}. With identical crust and mantle characteristics as the ones given in Table \ref{tab:Table2}, the filtered 4-layer models have an LVZ with a radius of $499_{497}^{501}$ km and a density of $3407_{3393}^{3413}$ kg/m$^3$. The LVZ viscosity ranges between 10$^{15.84}$ Pa$\cdot$s and 10$^{17.22}$ Pa$\cdot$s with a median value of 10$^{16.15}$ Pa$\cdot$s.
The core radius is $361_{343}^{391}$ km with a density of $5137_{4782}^{5451}$ kg/m$^3$. Its viscosity spreads over a wide range of values, from 10$^{17}$ to 10$^{26.60}$ Pa$\cdot$s, respectively for the 25$^{th}$ and 75$^{th}$ percentiles and a median value of 10$^{20.57}$ Pa$\cdot$s.
The $k-means$ clustering algorithm has identified two distinct sub-categories, so-called $Category\, 4_a$ and $Category\, 4_b$, gathering 864 models, namely 90\% of the models posterior to the $WRSS$ filtering (Table \ref{tab:Table3}). 

\begin{figure}
        \centering
            \includegraphics[scale=0.8]{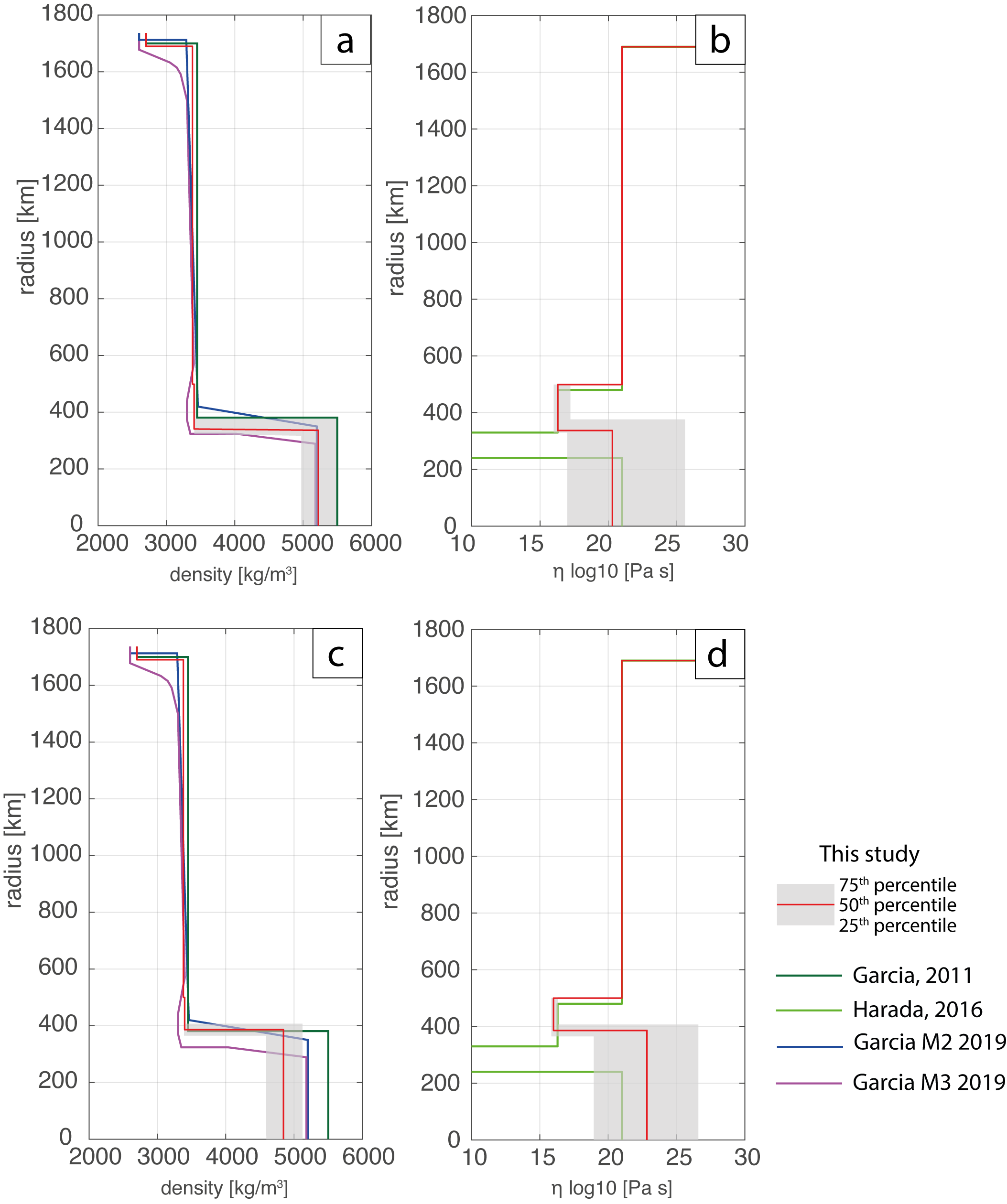}
                \caption{1-D density and viscosity ($\eta$) profiles of the $Category \,4_a$ (a-b) and $Category \,4_b$ (c-d) . The red lines correspond to the median of our sampling. The grey areas mark the range between the 25$^{th}$ and 75$^{th}$ percentiles. Colored lines show results from previous studies of \citet{2011PEPI..188...96G,harada2016deep,garcia2019lunar}.}
\label{fig:D-V-cat1}
\end{figure}

The 501 models of $Category\, 4_a$ have a thickness for the LVZ and the core of 154$\pm$24 km and 355$\pm$25 km, respectively (Table \ref{tab:Table3}). In contrast, the 363 models of $Category \, 4_b$ show a thinner LVZ and a thicker core of 101$\pm$23 km and 386$\pm$23 km, respectively (Table \ref{tab:Table3}).
Fig. \ref{fig:D-V-cat1} shows the density and the viscosity profiles deduced from our statistical approach for models of $Category\, 4_a$ (Figs. \ref{fig:D-V-cat1}-a,b) and $Category\, 4_b$ (Figs. \ref{fig:D-V-cat1}-c,d). As the crust and the mantle, for the most part, have constant model parameters, only the parameters of the LVZ and the core show a substantial variation. The median radius (500 km), density (3,405 kg/m$^3$) and viscosity (10$^{16.30}$ Pa$\cdot$s) for the models of $Category\, 4_a$ LVZ are close to that obtained for $Category\, 4_b$ (Table \ref{tab:Table4}). However, models of $Category\, 4_b$ have a less dense core compared to those models of $Category\, 4_a$. The viscosity of the core remains in the same order of magnitude for both categories. The eight orders of magnitude covered by the estimation of viscosity are due to a large dispersion of the models between the 25$^{th}$ and 75$^{th}$ percentiles. 
Compared to previous studies (\citet{2011PEPI..188...96G,weber2011seismic,garcia2019lunar}) the two categories fit in density, especially for the $Category\, 4_b$. The LVZ viscosity of the two groups fit with the 1-D profiles of \citet{harada2016deep}.

\begin{table}
        \centering
\caption{Internal Moon parameters for the $Category\,4$ models, after the $k-means$ clustering algorithm filtering. Values of parameters are the 50$^{th}$ percentile. Lower and upper scripts correspond to the 25$^{th}$ and the 75$^{th}$ percentiles.} 
            \begin{tabular}{c c c c c c c}
\hline
    Category & Layer & nb of models & Radius & Thickness & Density &  Viscosity \\
\hline
             & & k-means & km & km & kg/m$^3$ &  log10[Pa$\cdot$s] \\ 
\hline
\hline
    4$_a$ & LVZ & 501 & $500_{498}^{501}$ & $140_{133}^{159}$ & $3405_{3400}^{3413}$ & $16.30_{16}^{17.22}$ \\
    \\
    4$_a$ & Core & & $355_{341}^{376}$ & $355_{341}^{376}$ & $5223_{4973}^{5483}$ & $20.30_{17}^{25.60}$\\
    \\
\hline
    4$_b$ & LVZ & 363 & $499_{498}^{500}$ & $116_{92}^{124}$ & $3398_{3393}^{3413}$ & $16_{15.84}^{16.30}$\\
    \\
    4$_b$ & Core & & $386_{365}^{407}$ & $386_{365}^{407}$ & $4844_{4594}^{5123}$ & $22.84_{18.95}^{26.60}$\\
    \\
\hline
            \end{tabular}

\label{tab:Table4}
\end{table}


\subsection{5-layer ($Category \,5$)}
\label{sec:Category2}

The models of $Category \,5$ have five layers including an elastic inner core, an outer core and a LVZ (Table \ref{tab:Table2}). Over the 51,000 models, the WRSS filtering (Sect.\ref{sec:WRSS}) extracts 1,126 models (2.21\%) that agree with the observational constraints (Sect. \ref{sec:obs}). 
Similarly to $Category \,4$ models, the crust and mantle properties are constant as indicated in Table \ref{tab:Table2}. The selected models have an LVZ radius of $498_{498}^{500}$ km and a density of $3406_{3400}^{3413}$ kg/m$^3$.
The viscosity of the LVZ ranges between 10$^{14.30}$ Pa$\cdot$s and 10$^{23}$ Pa$\cdot$s with a median value of 10$^{16.60}$ Pa$\cdot$s. The outer core radius is $390_{363}^{421}$ km, with a thickness of $118_{67}^{170}$ km and a density of $4328_{4068}^{4711}$ kg/m$^3$. Its viscosity spreads over a wide range of values, from $10^{17.30}$ Pa$\cdot$s to $10^{23.73}$ Pa$\cdot$s, respectively for the 25$^{th}$ and 75$^{th}$ percentiles and a median value of 10$^{18}$ Pa$\cdot$s.

The radius of the inner core is $280_{233}^{310}$ km and its density is $6450_{5677}^{7553}$ kg/m$^3$. Among the 5-layer models, the $k-means$ clustering algorithm identified two distinct clusters of models gathering 562 models and representing 50$\%$ of the WRSS selections (see Table \ref{tab:Table3}), so-called $Category \,5_a$ and $Category\,5_b$ (Fig. \ref{fig:2Dhistob}). 
The second half of the models which are not considered by the $k-means$ clustering algorithm as potential clusters, cannot be ruled out and are kept as $Category\,5_c$. This group then gathers 564 models and their main characteristics are presented in Table \ref{tab:Table5}.

The 320 models of $Category \,5_a$ shows thicknesses for the LVZ and the outer core of 123$\pm$22 km and 76$\pm$14 km respectively with an inner core radius of 304$\pm$26 km. In contrast, the 242 models of $Category \,5_b$ reveal close thicknesses for the outer core and LVZ, of about 123 km for the LVZ and 147 km for the outer core. However, the inner core is thinner than the one of the $Category \,5_a$ with a radius of 226$\pm$23 km to be compared with the 304$\pm$20 km for $5_a$. In addition, the $Category \,5_b$ outer core is much thicker (140 km) than the one of $Category \,5_a$ (70 km) but with a two-fold dispersion for the $Category\,5_b$ with respect to the $Category\,5_a$. Finally, the outer core viscosity of models of $Category \,5_a$ is more than two orders of magnitude lower than that of models of $Category \,5_b$ with an equivalent dispersion for both sub-categories.

As expected by its construction, $Category\,5_c$ shows a bigger dispersion of the outer core radii and viscosities than $Category \,5_a$ and $Category \,5_b$: the dispersion for the inner core radius is for example 2.3 times the one of  $Category \,5_b$ and 1.95 the one of $Category \,5_a$. On the other hand, LVZ thickness and viscosity for $Category\,5_c$ are almost as accurate as the two other categories. 
It is also interesting to note that the $Category\,5_c$  values for the inner and outer cores appear to be in between the estimations of $Category \,5_a$ and $Category \,5_b$ with a radius for the inner core of about 277 km when the one of the $Category \,5_b$ is 18$\%$ smaller and the one of $Category \,5_a$ is 9$\%$ larger. The same holds for the thickness of the outer core with a value of 102 km for $Category \,5_c$ when the $Category \,5_b$ gives an outer core thickness 44$\%$ larger and $Category \,5_a$, 25$\%$ smaller. 

For the LVZ, the mechanism seems to be different. The $5_c$ estimations for the LVZ thickness are close to one of the other two sub-categories with a dispersion (2 km) equivalent or smaller than the one of $Category \,5_a$ (2 km) and $Category \,5_b$ (3 km). An important remark stands for the outer core and LVZ $5_c$ viscosities. Contrary to the previous two sub-categories favoring an LVZ less viscous than the outer core, $Category \,5_c$ gives an LVZ almost 8 orders of magnitude more viscous than $Category \,5_a$ and $5_b$ for an outer core one order of magnitude less viscous than $5_a$ and almost 4 orders of magnitude less than $5_b$. We note that the dispersion for the outer core viscosity for $Category \,5_c$ is smaller than the one of $Category \,5_a$ and $Category \,5_b$, both $Category \,5_a$ and $Category \,5_b$ viscosity values encompassing this latest one.
With $Category \,5_c$, we then have a new type of lunar interior profile with an outer core less viscous than the LVZ. We can also stress the important dispersion of the  $Category \,5_c$ LVZ viscosity (of about 7 orders of magnitude compared to only 2 orders of magnitude for $5_a$), leaving room for even overlap of values between the outer core and LVZ viscosities. 

\begin{figure}

        \centering
                \includegraphics[scale=0.7]{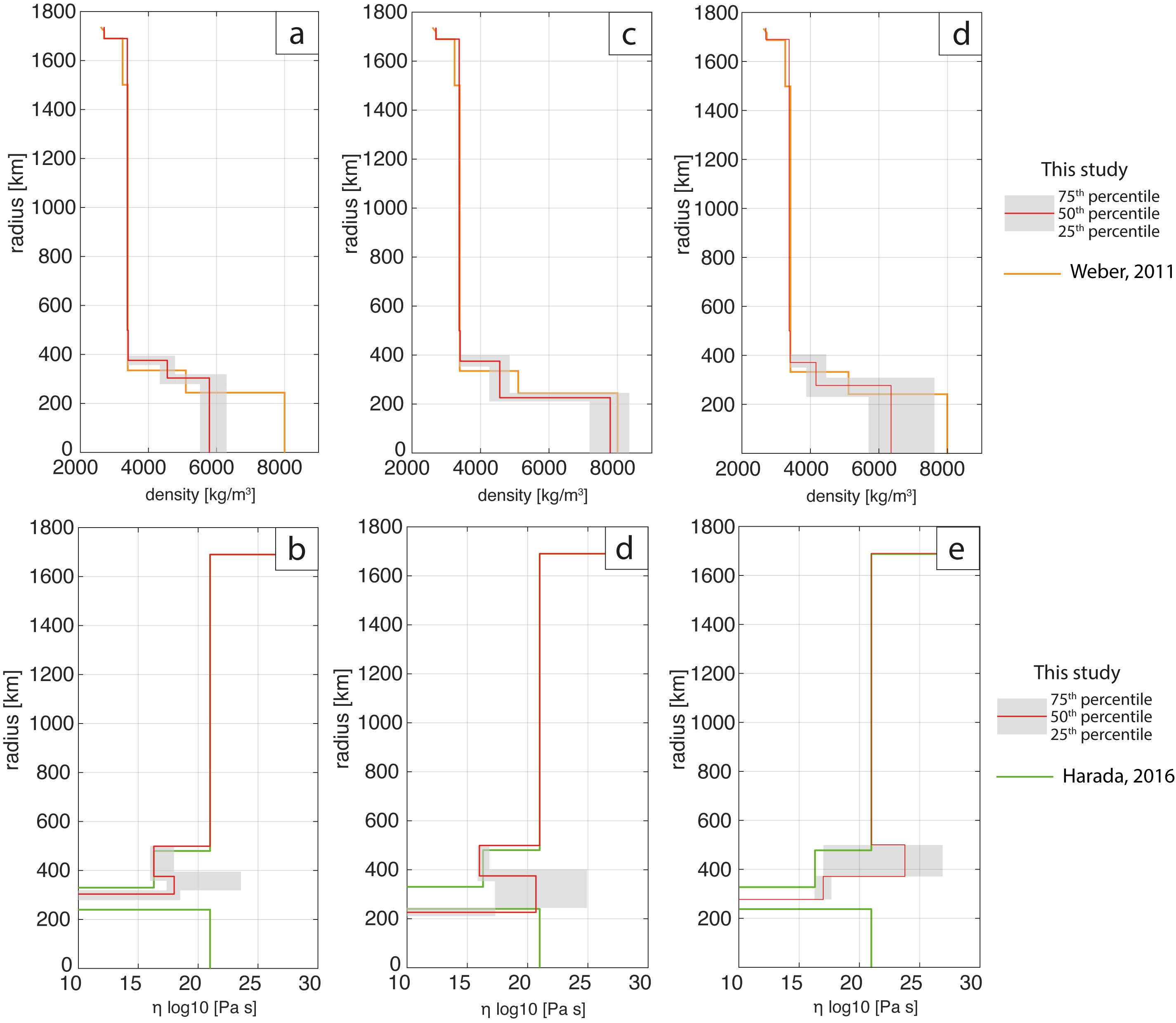}
                \caption{1-D density and viscosity ($\eta$) profiles for the two sub-categories deduced from the 5-layer \af{modeling}: $Category \,5_a$ (a-b),  $Category \,5_b$ (c-d) and $Category \,5_c$ (e-f). The red lines correspond to the median of our sampling. The grey area is the 25$^{th}$ and 75$^{th}$ percentiles. Colored solid lines are previous studies of \citet{ weber2011seismic,harada2016deep}.}
    \label{fig:D-V-cat2}
\end{figure}

Fig. \ref{fig:D-V-cat2} shows the 1-D profiles deduced from our statistical approach for $Category\, 5_a$ (Figs. \ref{fig:D-V-cat2}-a,b), $Category\, 5_b$ (Figs. \ref{fig:D-V-cat2}-c,d) and $Category\, 5_c$ (Fig{s}. \ref{fig:D-V-cat2}-e,f). Table \ref{tab:Table5} lists the parameters of the Moon internal structure deduced for the models of the three sub-categories. We retrieve for the LVZ very similar radius, density and viscosity as those obtained for models of $Category \,4_a$ and $Category \,4_b$ (Sect. \ref{sec:Category1}). 
The three viscosities of the outer and inner cores diverge from the study of \citet{harada2014strong, harada2016deep} due to the rheologies used in our study. Indeed, \citet{harada2014strong} assume that the outer core is an in-viscid fluid while we assume a Newtonian outer core. We also consider a purely elastic inner core, meaning that its viscosity tends to infinity. This assumption prevents us to estimate a viscosity for the inner core.

\begin{table}
        \centering

\caption{Lunar interior parameters after the $k-means$ clustering algorithm filtering for the 5-layer \af{modeling}. {Confidence intervals are reported with the same notation used in} Table \ref{tab:Table4}.} 
            \begin{tabular}{c c c c c c c}
\hline
    Category & Layer & nb of models & Radius & Thickness & Density &  Viscosity \\
    \hline
             & & k-means & km & km & kg/m$^3$ & log10[Pa$\cdot$s] \\ 
\hline
\hline
    5$_a$ & LVZ & 320 & $499_{498}^{500}$ & $123_{104}^{156}$ & $3406_{3400}^{3413}$ & $16.30_{16}^{17.98}$ \\
    \\
    5$_a$ & Outer core & & $376_{358}^{395}$& $76_{67}^{90}$ & $4065_{3918}^{4267}$ & $18_{17.37}^{23.57}$\\
    \\
    5$_a$ & Inner core & &  $304_{279}^{319}$ & $304_{279}^{319}$ & $5789_{5517}^{6296}$ & -- \\
    \\
   \hline
    5$_b$ & LVZ & 242 & $499_{497}^{500}$ & $123_{98}^{146}$& $3407_{3400}^{3413}$ & $16_{15.90}^{16.84}$\\
    \\
    5$_b$ & Outer core & & $375_{353}^{402}$ & $147_{125}^{169}$ & $4276_{4041}^{4483}$ & $20.69_{17.30}^{24.95}$\\
    \\
    5$_b$ & Inner core & &  $226_{211}^{245}$ &$226_{211}^{245}$ & $7787_{7183}^{8341}$ & --\\
    \\
   \hline
    5$_c$ & LVZ & 564 & $500_{499}^{501}$ & $127_{94}^{148}$ & $3406_{3400}^{3413}$ & $23.77_{16.95}^{26.90}$\\
    \\
    5$_c$ & Outer core & & $371_{350}^{405}$ & $102_{79}^{130}$ & $4163_{3881}^{4461}$ & $17_{16.77}^{17.69}$\\
    \\
    5$_c$ & Inner core & &  $277_{231}^{309}$ & $277_{231}^{309}$ & $6353_{5695}^{7619}$ & --\\
    \\
\hline
            \end{tabular}
\label{tab:Table5}
\end{table}

\section{Considerations regarding LVZ temperature and consequences}
\label{sec:part5}

In the previous sections, we presented a selection of models for the lunar interior satisfying within $3$-$\sigma$ uncertainties of the observational constraints on frequency-dependent dissipation terms and TLNs derived from LLR and GRAIL. From this selection, we showed the distribution of geophysical parameters describing the lunar internal structure.

From our results presented in Sect. \ref{sec:prem_results}, the LVZ radius and density are well constrained. Conversely, the viscosity of the LVZ varies by about two orders of magnitude within the range of quantiles (see Tables \ref{tab:Table4} and \ref{tab:Table5}). However, our estimate of the median of the LVZ viscosity agrees with previous findings (\citet{harada2014strong,tan2021tidal}) of about $10^{16}$ Pa$\cdot$s to $10^{17.60}$ Pa$\cdot$s. The LVZ radius and densities are in agreement with previous studies of \citet{weber2011seismic,harada2014strong,harada2016deep}, namely about 500 km and 3,400 kg/m$^3$, respectively.

At this point, it is reasonable to question what we can infer about the status of the LVZ from our statistical modeling approach. As it was stated in seismological studies, the LVZ is supposed to be a part of the mantle with high viscosity (at about $10^{21}$ Pa$\cdot$s), meaning that this thin layer should have crossed the lunar mantle solidus for justifying such low viscosity profiles (between $10^{15.90}$ - $10^{16.95}$ Pa$\cdot$s). In this sense, it is worth to investigate the temperature profiles of our selected models for the LVZ since we obtained very stringent constraints for this layer.

\subsection{On constraining the lunar deep mantle temperature}
    \label{sec:temperature_lvz}
    
The $Categories$ $4$ and $5$ display a wide range of viscosities considering the quantiles as listed in Tables \ref{tab:Table4} and \ref{tab:Table5}. 
The LVZ of the lunar mantle controls the seleno-dynamic processes (\citet{harada2014strong, harada2016deep}), so to better constrain the lunar models, we use the mantle and LVZ as constraining layers.

Along the lines of \citet{nakada2012viscosity}, who relate the depth-varying viscosity {$\eta(z)$} to the temperature of the lunar mantle, we assume that the LVZ viscosity depends upon temperature as follows{:}
\begin{equation}
    \label{eq: temp1}
        \eta(z) = \eta_0 \exp\left(\frac{H^*}{R_g T}\right)
\end{equation}
where {$\eta_0$=10$^{21}$ Pa$\cdot$s is the mantle viscosity,} $H^*$ is the activation enthalpy and $R_g$ is the gas constant.

The depth of the upper and lower boundaries of the LVZ (\emph{i.e.} the mantle-LVZ interface and the LVZ-core interface) are defined as $z_m$ and $z_{lvz}$, respectively. These quantities are related to the thickness of each category given in Sect. \ref{sec:Category1} and \ref{sec:Category2}. Here, we assume that at $z=z_m$ the viscosity is equal to the mantle viscosity, \emph{i.e.} $\eta(z_m)=\eta_m$ (see Table. \ref{tab:Table2}), while the temperature at {the mantle-LVZ interface}, $T(z_m)=T_m$, varies according to the unknown depth of the top LVZ boundary.

We take the temperature {at the mantle-LVZ interface} ($T_m$) as defined by \citet{khan2014geophysical}. In their work, \citet{khan2014geophysical} have performed marginals-posterior probability density function (\emph{i.e.} PDF) profiles of the Moon depicting the modeled temperature as a function of depth. They have fixed depth node histograms, reflecting the PDF of the sampled temperatures. By lining up these marginals, the temperature can be envisioned as contours directly related to probability occurrence. The range of admitted LVZ temperatures is listed in Table \ref{tab:tab_temp}.

Following \citet{nakada2012viscosity}, by setting $\eta(z_m)=\eta_m$ we rewrite Eq. (\ref{eq: temp1}) as:
\begin{equation}
    \label{eq: temp1_1}
        \eta(z) = \eta_m \exp\left[-\frac{H^*}{R_g} \left(\frac{1}{T_m} - \frac{1}{T(z)}\right)\right].
\end{equation}
We then assume that the temperature within the LVZ scales with depth as $T(z) = T_m +  \Delta T(z)$, hence the temperature at the LVZ-core interface is $T(z_{lvz}) = T_{lvz} = T_m + \Delta T_{lvz}$. In analogy with \citet{nakada2012viscosity}, from Eq. (\ref{eq: temp1_1}) we obtain:
\begin{equation}
    \label{eq: temp2}
        \eta_{lvz} = \eta_m \exp \left(-\frac{H^{*}}{R_g T_m} \frac{\Delta T_{lvz} / T_m}{1+\Delta T_{lvz} / T_m}\right)
\end{equation}
or, equivalently:
\begin{equation}
    \label{eq: temp3}
        \ln\frac{\eta_{lvz}}{\eta_m} = \frac{H^{*}}{R_g T_m} \frac{\Delta T_{lvz} / T_m}{1+ \Delta T_{lvz} / T_m}
\end{equation}
which provides the temperature ratio $\Delta T_{lvz}/T_m$ as a function of the viscosity ratio $\eta_{lvz}/\eta_m$:
\begin{equation}
    \label{eq: temp4}
        \frac{\Delta T_{lvz}}{T_m} = \frac{\ln\left(\frac{\eta_{lvz}}{\eta_m}\right)}{\frac{H^*}{R_g T_m} - \ln\left(\frac{\eta_{lvz}}{\eta_m}\right)}.
\end{equation}
\af{From} Eq. (\ref{eq: temp4}), the temperature at the LVZ-core interface can be obtained as a function of two unknowns, \emph{i.e.} the temperature at the mantle-LVZ interface ($T_m$) and the activation enthalpy ($H^*$). The $\eta_{lvz}/\eta_m$ ratio is estimated from our results for each category (Tables \ref{tab:Table4},\ref{tab:Table5}).

As highlighted in Sect. \ref{sec:part5}, the LVZ thickness is well constrained by our statistical modeling. We take then a depth of 1237 $\pm$ 2 km for defining $z_{m}$. According to \citet{khan2006earth} the posterior temperature profile at this depth correspond to the range [1200°C - 1500°C], depending on the least and most probable occurrences. 
The second unknown is $H^*$. On one hand, we assume that the lunar mantle is "a\, priori" 
composed as  the Earth upper mantle (\citet{katz2003new,tomlinson2021thermodynamic}). We adopt a peridotite composition, which is composed of variable proportions of olivine, orthopyroxene, clinopyroxene and aluminium phase (garnet/spinel). From the meta-stable phases of minerals, depending on the water content, the activation enthalpy ($H^*$) must be encompassed in between [372:430] kJ/mol under 4.5 GPa (or 1237 km depth, \citet{nakakuki2010dynamical,yamazaki2001some}). On the other hand, Ilmenite-bearing cumulates enriched with TiO$_2$ may favor the existence of a partially molten layer at the lunar core-mantle boundary. Hence, we take into account the $H^*$ of the Illmenite at about [275:283] kJ/mol \citep{tokle2021effect}. Table \ref{tab:tab_temp} summarizes the parameters used for computing the Moon mantle temperature. With such modeling, we are able to compare the temperature profiles for the LVZ and the mantle to the Earth solidus hypothesis found in the literature.

\begin{table}[ht]
        \centering

\caption{Parameters used for constraining the LVZ temperature. [1] \citep{yamazaki2001some}, [2] \citep{tokle2021effect},[3] \citep{nakakuki2010dynamical}, [4] \citep{khan2004does}, [5] \citep{khan2014geophysical}} 
            \begin{tabular}{c c c c}
            Symbol & Value & Unit & Reference \\
            \hline
            \hline
            {$R_g$} & 8.314462 & J$\cdot$ K$^{-1}$ $\cdot$ mol$^{-1}$ & -- \\
            {$H^*$} &  {275 - 430} & kJ/mol & [1],[2],[3]\\
            {$z_m$} & 1237 $\pm$ 2 & km  & this study \\
            {$\eta_m$} & 1$\times$10$^{21}$ & Pa$\cdot$s & this study\\
            {$\eta_{lvz}$} & 8$\times$10$^{15}$ - 9$\times$10$^{17}$ & Pa$\cdot$s & this study\\
            {$T_m$}  & 1200 - 1500 & °C & [4],[5] \\
             \hline
\end{tabular}
\label{tab:tab_temp}
\end{table}

\subsection{Use of temperature profile}
\label{sec:temp_profile_use}

\begin{figure}
\centering
        \centering
        \includegraphics[scale=0.7]{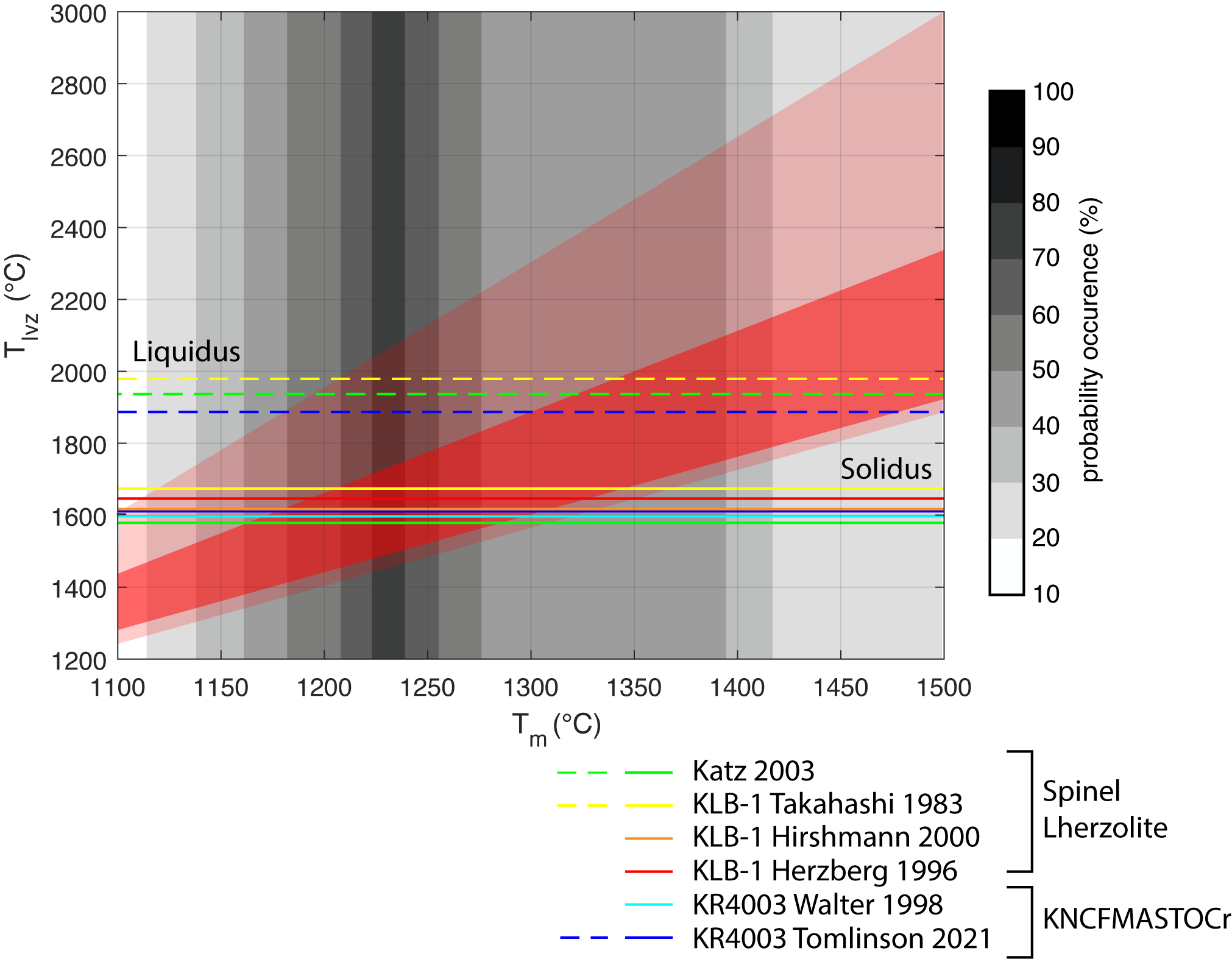}
        \caption{LVZ temperature ($T_{lvz}$) as function of mantle temperature ($T_m$). The shaded red area corresponds to the minimum and maximum LVZ viscosity found in this study for both Categories. The red area marks the median of LVZ viscosity. Variations correspond to the range of $H^*$ at 1237 km depth. Grey areas are the probability occurrence of mantle-LVZ interface temperature at the depth of 1237 km from \citet{khan2004does}. Colored solid and dashed lines are the solidus and liquidus from \citet{takahashi1983melting,herzberg1996melting,walter1998melting,hirschmann2000mantle,katz2003new,tomlinson2021thermodynamic}.}
\label{fig:temp_fig}
\end{figure}

With the approach described in Sect.\ref{sec:temperature_lvz}, we attempt to better constrain the lunar interior characteristics.
Fig. \ref{fig:temp_fig} summarises the relations between the admitted range of mantle temperature ($T_m$) at the mantle-LVZ interface from \citet{khan2006earth} and the temperature at the LVZ-core interface ($T_{lvz}$), deduced from Eq. (\ref{eq: temp4}). The shaded red area corresponds to the range of LVZ viscosities (i.e., $1^{st}$ and $3^{rd}$ quantiles) while the red area indicates the median of both categories. The dispersion around the median is due to the variation in $H^*$ linked to its variability, depending on the mineral phase (i.e., peridotite or ilmenite). The vertical grey areas correspond to the probability occurrence of temperature at the given depth of 1237$\pm$2 km modelled by \citet{khan2001new}. The solidus and liquidus deriving from studies of \citet{takahashi1983melting, herzberg1996melting,walter1998melting,hirschmann2000mantle,katz2003new,tomlinson2021thermodynamic} correspond to the olivine assemblage at depth $z_m$.
Here, our range of LVZ viscosity cross-cuts the solidus in between 1560°C and 1720°C within the area of 70\% probability occurrence of mantle-LVZ interface temperature ($T_m$). From this constraint in temperature we can deduce a narrow range of the LVZ viscosity of {10$^{16.3}$} Pa$\cdot$s and {10$^{18}$} Pa$\cdot$s.
The low viscosity might be in favor of possible partial melting of the LVZ as previously suggested in \citet{weber2011seismic,khan2014geophysical,tan2021tidal}.
We use Fig. \ref{fig:temp_fig} for further filtering our selection of models from the Sect. \ref{sec:prem_results}. This selection of models correspond to 5.5\% of the statistically selected models and their main characteristics are presented in Table \ref{tab:Table6} and Figs. \ref{fig:4a} and \ref{fig:5a}.

\section{Discussion}\label{sec:discussion}

In Sect. \ref{sec:prem_results} we present statistical selections obtained considering the total mass and the MoI of the Moon and the WRSS as selection criteria. 
We have seen that five categories of models were obtained: three including a solid inner core and an outer core together with a crust, a mantle and an LVZ and two with only one uniform core. On the basis of the present observations, we are not able to support or reject the hypothesis of the existence of a Moon inner core. Another important result of the Sect. \ref{sec:temperature_lvz} is the consistent constraint obtained on the LVZ thickness, which turns out to be common to the five selected categories as well as for its viscosity, common for four out of five categories.
Figs. \ref{fig:1Dprofiles}, \ref{fig:D-V-cat1}, \ref{fig:D-V-cat2} confirm that our five possible series of models are consistent with the profiles deduced from seismological data such as \citet{weber2011seismic,2011PEPI..188...96G,garcia2019lunar}.
In Sect. \ref{sec:temp_profile_use} we introduce an additional selection criterion by considering only models with the LVZ temperature profiles consistent with an intersection of the solidus at the mantle-LVZ depth and temperature (Fig. \ref{fig:temp_fig}). 

\begin{figure}[ht]
\centering
\includegraphics[scale=0.4]{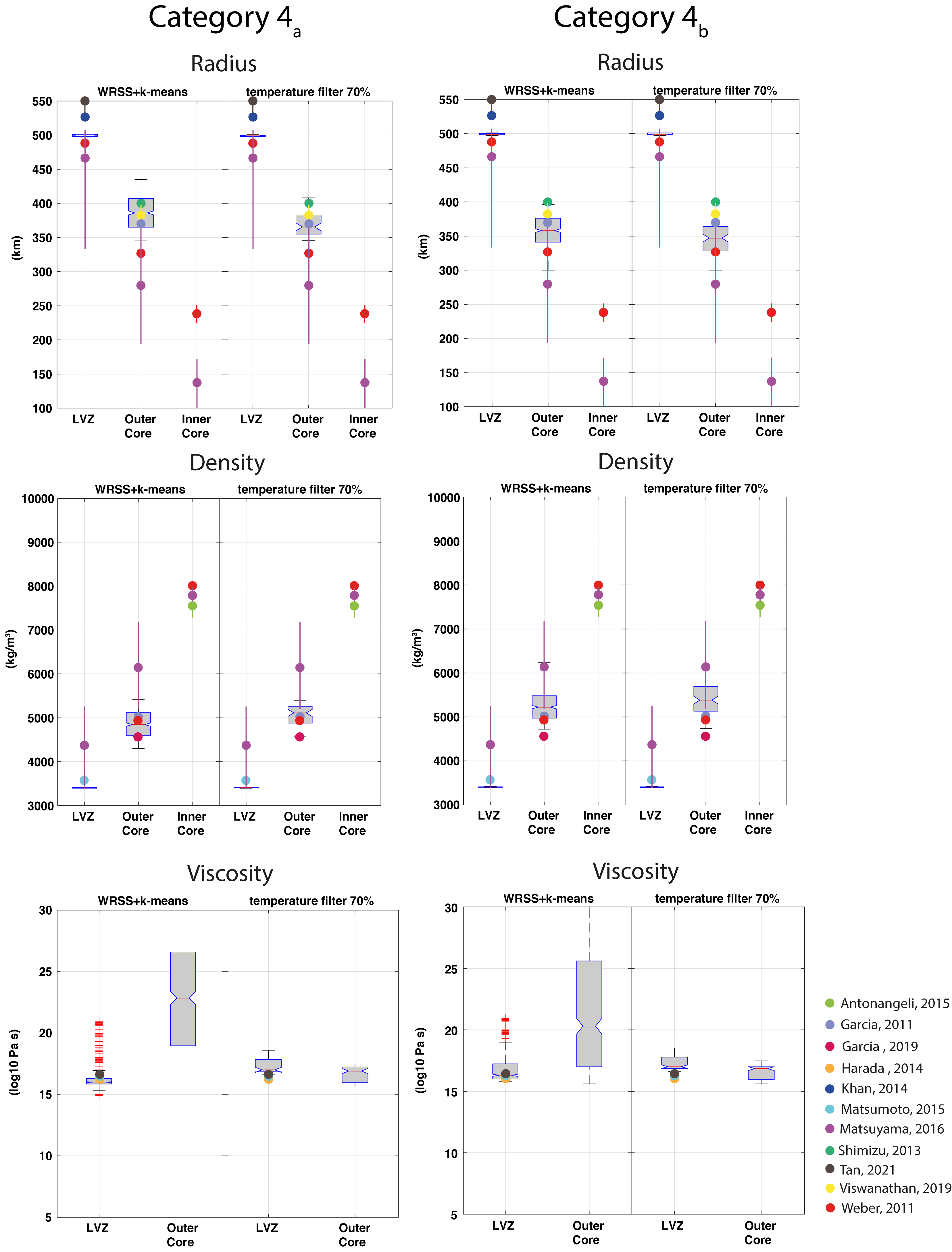}
        \caption{Distributions of radius, density and viscosity for each layer in models of $Category\,4_a$ (left panels) and $4_b$ (right panels). Colored dots mark results from previous studies. \af{Are also represented measures obtained from non-geodesic techniques such as the magnetic soundings from \cite{2013Icar..222...32S}}}
\label{fig:4a}
\end{figure}

\begin{figure}[ht]    
\centering
\includegraphics[scale=0.4]{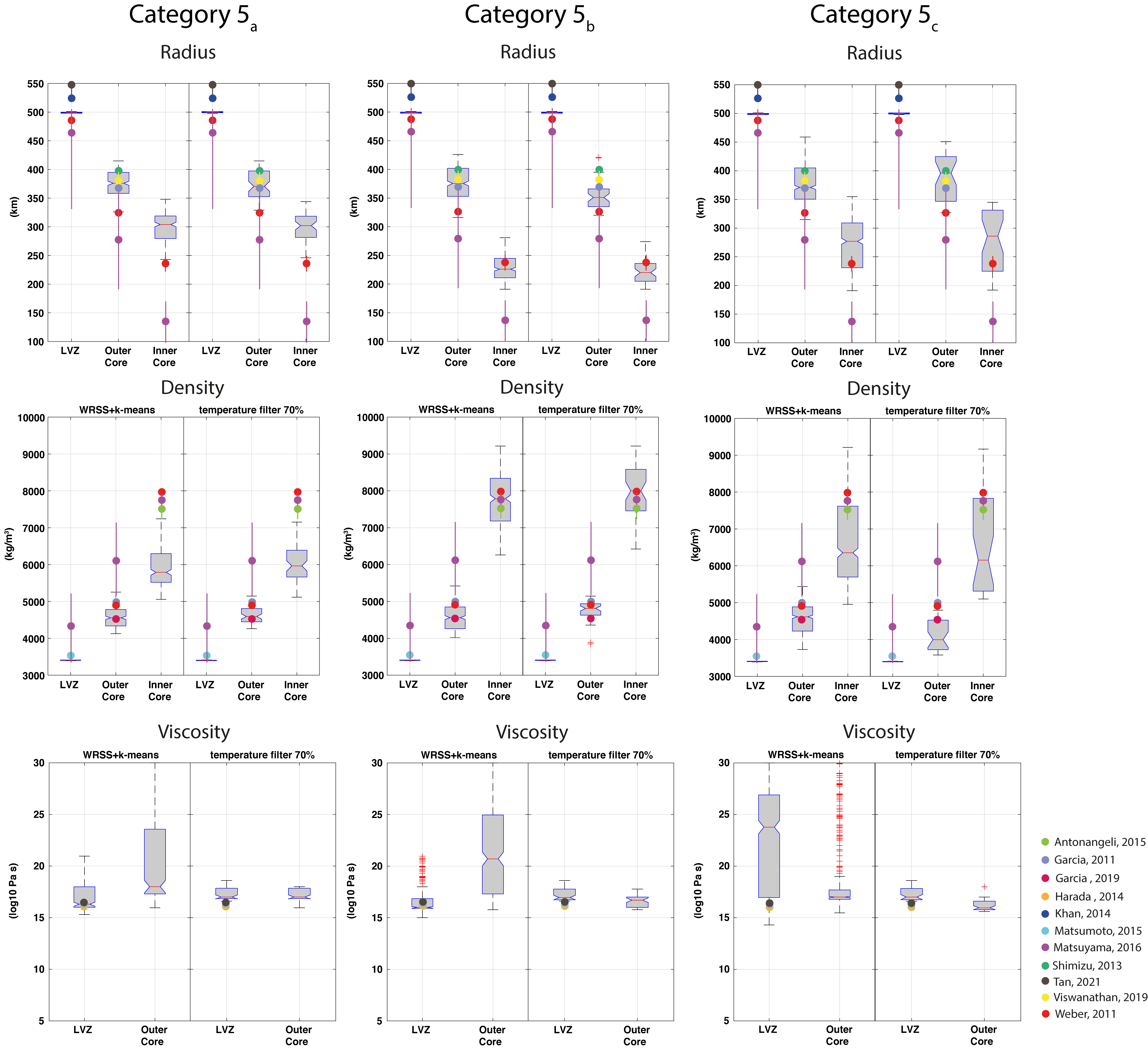}
        \caption{Distributions of radius, density and viscosity for each layer in models of $Category\,5_a$ (left panels), $5_b$ (middle panels) and $5_c$ (right panels). Colored dots mark results from previous studies. \af{Are also represented measures obtained from non-geodesic techniques such as the magnetic soundings from \cite{2013Icar..222...32S}}}
\label{fig:5a}
\end{figure}

In considering Fig. \ref{fig:temp_fig}, we are able to keep only a sub-sample of the $Categories$ $4$ and $5$. The main characteristics of the remaining models are provided in Table \ref{tab:Table6} and in Figs. \ref{fig:4a} and \ref{fig:5a} in which results obtained with statistical filtering described in Sect. \ref{sec:prem_results} are plotted next to these new results.
It is important to note that the intersection between our model groups and the solidus obtained for different mantle chemical compositions match the range of temperatures (between 1600 and 1800 °C) expected for the Earth mantle below 4.5 GPa. This is a good indicator of the consistency of our results. 
An important result is the significant reduction of the dispersion in the distribution of the LVZ and outer core viscosities of $Categorie$ $5_b$ and $5_c$ after the application of the temperature filter. This is also true for all five categories of models, either including or not an inner core. On average, when for the WRSS+$k$-means filtering the dispersion between the first and the last quantiles of the outer core viscosity was of about 8 orders to magnitude, it is less than 2 orders of magnitude with the temperature filter. This allows the first accurate determination for the outer core viscosity of about {16.90$_{15.95}^{17.82}$} Pa$\cdot$s without inner core and of {15.95$_{15.77}^{17.84}$} Pa$\cdot$s with inner core.

Besides these results, it is also interesting to note that our estimations are consistent with previous studies as one can see in Figs. \ref{fig:4a} and \ref{fig:5a}. Considering the 5-layer modeling, the $Category\,5_b$ seems to be more in accordance with the previous studies with a good match of the inner core density and size, especially with \citet{weber2011seismic}. The $Category \,5_b$ is also consistent with \citet{garcia2019lunar} regarding the density of the outer core. Considering the LVZ, our estimations meet the error bars from \citet{2016GeoRL..43.8365M} in density and radius and match well with \citet{tan2021tidal} for the viscosity. The $Category\,5_a$ proposes an alternative series of models that can be ruled out by considering our statistical or temperature filterings. These models propose a smaller and less dense inner core. Instead of an inner core of about 220 km (and a density of 8000kg/m$^{3}$) deduced with the models of $Category\,5_b$, the inner cores of $Category\,5_a$  have a radius of about 302 km with a density of 5830 kg/m$^{3}$. The $Category\,5_a$ might correspond to a new series of models with a less metal-rich inner core component in comparison with the $Category\,5_b$ models which are in favor of a Fe inner core. Less dense inner cores may favour the presence of volatile-rich elements according to the core differentiation models \citep{steenstra2017lunar}. A large inner core (i., $Category\,5_a$ and $5_c$) would be indeed, enriched in light elements and resolving thereby, the so-called core density deficit (CDD) because the resulting alloy would have an expanded volume and reduced average atomic mass relative to pure iron \citep{khan2018geophysical,stahler2021seismic,murphy2016hydrogen}. Nevertheless, since our models are based on geodetic constraints, further geochemical analysis would be required to explore the reliability of the core density with a volatile-rich composition. Regarding the Newtonian outer core and the LVZ, the $Category\,5_a$ and 5$_b$ give very close results.
The $Category \, 5_c$ proposes an intermediate value for the outer core thickness (of about 133 km) but an inner core smaller than in $Category \, 5_a$. The estimations for the outer core viscosities are, for the three categories, inside the quantiles intervals with a larger dispersion for $Category \, 5_c$. In considering the largest interval of dispersion over the three sub-categories, we end up with the outer core viscosity for the five-layer modeling of about {16.54$_{15.77}^{17.84}$} Pa$\cdot$s.
For the LVZ, the $Category \, 5_c$ gives the thinnest value with 102 km against 148 km for  5$_b$ and 132 km for 5$_a$. The LVZ viscosities are very similar for the three categories with a dispersion of less than $10^{1.5}$ Pa$\cdot$s. 
Finally, while, before the temperature filtering, the $Category \, 5_c$ gathers about the same number of selected models than $Category \, 5_a$ and 5$_b$ together, only 35 $5_c$ models pass the solidus line against 65 for $Categories \, 5_a$ and 50 for 5$_b$. Besides showing the importance of the temperature filtering for our final results, these figures seem also to indicate that the models favoring a more viscous LVZ than the outer core do not meet easily the constraint of a melting LVZ between 1600 and 1800°C.

For the four-layer modeling, the differences between the $Category \, 4_a$ and 4$_b$ are less important than between the $Category \, 5_a$ and 5$_b$. Mainly one can notice that the temperature filtering induces a reduction of the dispersion of the outer core thicknesses which is more pronounced in $Category \, 4_a$ than in 4$_b$. For $Category \, 4_a$, the use of the temperature profiles as a filter induces a reduction of about 50$\%$ of the dispersion while the dispersion for 4$_b$ seems to be unaffected by the temperature filtering. In terms of consistency with the previous studies, our results of 4$_a$ and 4$_b$ are in the range of the values found in the literature. One can notice that the temperature profile selection tends to favor the \citet{2011PEPI..188...96G} relative to \citet{viswanathan2019observational} and \citet{Antonangeli3916} regarding the radius of the core. In addition, the $Category\,4$ suggests a density for a fully molten in the range of proposed densities at the pressure and temperature (T $\gg$ 1800K) of the lunar core. These ranges of densities can be explained by liquid Fe-S or Fe-Ni-Si alloys \af{containing 30 to 40\% in weight of} sulfide content \citep{morard2018liquid,terasaki2019pressure}.

\begin{table}[ht]
        \centering
        \caption{Characteristics of categories layers using the temperature filtering (see Sect. \ref{sec:part5} for more details.}
            \begin{tabular}{l c c c c c c c}
\hline
    Category & nb of models & Layer & Radius & Thickness & Density &  Viscosity \\
    \hline
             & T°C filtering & &  km & km & kg/m$^3$ &  log10[Pa$\cdot$s] \\ 
\hline
\hline
    4$_a$ & 58 & LVZ & $499_{498}^{500}$  & $132_{117}^{143}$ & $3402_{3400}^{3413}$ &
    $17.00_{16.84}^{17.82}$ \\
    4$_a$ & & {Core} & $366_{355}^{383}$  & $366_{355}^{383}$ & $5111_{4878}^{5262}$ &
    $16.90_{15.95}^{17.00}$ \\
    \hline
    4$_b$ & {99} & {LVZ} & $499_{498}^{501}$ & $140_{123}^{159}$ & $3406_{3400}^{3413}$ & $17.00_{16.85}^{17.77}$ \\
    \\
    4$_b$ &  & {Core} & $347_{328}^{363}$ & $347_{328}^{363}$ & $5383_{5131}^{5690}$ & $16.84_{15.96}^{17.00}$\\
    \hline
    5$_a$ & {64} & {LVZ} & $500_{499}^{501}$  & $129_{101}^{147}$ & $3400_{3393}^{3406}$  &  $16.84_{16.30}^{17.84}$ \\
    \\
    5$_a$ & & {Outer core} & $370_{352}^{397}$ & $76_{67}^{85}$ & $4272_{3987}^{4423}$ & $17.00_{16.80}^{17.84}$ \\
    \\ 
    5$_a$ & & {Inner core} & $302_{281}^{318}$ &  $302_{281}^{318}$ & $5830_{5525}^{6276}$ & $--$ \\
    \hline
    5$_b$ & {50} & {LVZ} & $499_{498}^{500}$ & $148_{132}^{164}$ & $3407_{3400}^{3413}$ &  $16.95_{16.77}^{17.77}$ \\ %
    \\
    5$_b$ & & {Outer core} & $351_{335}^{366}$ & $133_{110}^{148}$ & $4537_{4355}^{4666}$  & $16.69_{16.00}^{17.00}$ \\ 
    \\
    5$_b$ & & {Inner core} & $220_{205}^{236}$ & $220_{205}^{236}$ & $8000_{7457}^{8585}$  & $--$ \\
        \hline
    5$_c$ & {35} & {LVZ} & $500_{499}^{501}$ & $102_{75}^{151}$& $3413_{3398}^{3420}$ &  $16.85_{16.77}^{16.82}$ \\ %
    \\
    5$_c$ & & {Outer core} & $396_{346}^{445}$ & $113_{84}^{133}$ & $3996_{3723}^{4525}$  & $15.95_{15.77}^{16.60}$ \\ 
    \\
    5$_c$ & & {Inner core} & $286_{224}^{332}$ & $286_{224}^{332}$ & $6153_{5313}^{7832}$  & $--$ \\
\hline
            \end{tabular}
            
\label{tab:Table6}
\end{table}
\clearpage
\section{Conclusions}
\label{sec:conclusions}

In this work, we presented a selection of possible modeling for the Moon structures on the basis of observational constraints on tidal deformation and dissipation.  We adapted the semi-analytical code $ALMA$, originally aimed at evaluating time-domain LNs suitable for the Earth, to the estimation of frequency-dependent tidal LNs for a given lunar interior model.
We generated 120,000 random models of the lunar interior in which the thicknesses and viscosity profiles are varied within plausible ranges and the mantle and crust parameters are kept constant while keeping the mass, MoI and seismological profiles consistent with current determinations. For each model, we estimated tidal LNs and dissipation coefficient at two periods. We selected 1462 models that fit with the present observational constraints. As this selection of models provides very accurate information on the LVZ thickness and viscosity, confirming the viscosity gradient between the upper mantle and the core-mantle boundary, we further refine our ensemble of models by requiring their temperature profiles to be consistent with the hypothesis of an intersection with the solidus, as seen for the Earth mantle. Our findings can be summarized as follows:

\begin{enumerate}
\item The current selenodetic constraints (i.e., the mass of the Moon, MoI, TLNs, dissipation coefficient and seismic velocity) cannot clearly rule out the presence of an inner core.  
\item On the basis of our geodetic statistical filtering, we can conclude that the LVZ is well constrained with a radius of (500 $\pm$ 1) km, a density of (3400 $\pm$ 10) kg/m$^3$ and a viscosity of about {17.00$_{16.84}^{17.82}$} Pa$\cdot$s without inner core and {16.88$_{16.30}^{17.84}$} Pa$\cdot$s with inner core. Both estimations are consistent within the quantile intervals.
\item We obtain the first estimation for the viscosity of the outer core. The viscosity of the core is of about 16.87$_{15.95}^{17.00}$ Pa$\cdot$s without inner core and of 16.54$_{15.77}^{17.84}$ Pa$\cdot$s with an inner core.
\end{enumerate}

Besides these main results, one can also stress two possible scenarios regarding an inner core.
\begin{itemize}
    \item One category (i.e., 5$_a$) of models favors a big inner core of about 302 km radius but with a density smaller than the one expected for a pure iron core (about 6000 kg/m$^{3}$) and a small outer core with a thickness of about 76 km and a density of 4280 kg/m$^{3}$. These models are consistent with a less dense metal-rich inner core as previously thought \citep{murphy2016hydrogen,khan2018geophysical,stahler2021seismic}. The category 5$_c$ shows wider dispersion in radius and densities. Only the upper bound of the density interval may suggest an iron-rich inner core while the lower bound may correspond to a new type of composed iron-light elements alloys. Less dense inner cores may favour the presence of volatile-rich elements according to the core differentiation models of \citet{steenstra2017lunar}. However, the models presented in this study are based on geodetic constraints and statistical analysis only. Further geochemical analysis in laboratory would be required to further explore the reliability of the core density with volatile-rich composition.
    \item The other category is closer to the traditional picture of the telluric planet model with a dense inner core, about 8000kg/m$^{3}$ for a 220 km radius, and a thicker outer core with a thickness of about 133 km.
\end{itemize}

 The investigation on plausible temperature dependency of the viscosity of the LVZ gives insights into the partially molten conditions as well as the thermal state of the Moon mantle. As suggested in \citet{harada2016deep} the LVZ might play the role of thermal blanket for the cooling on the core which might result in degree-one convection and explains the formation of lunar mare basalts asymmetry (\citet{zhong2000dynamic}).

\break
\section*{Acknowledgements}
This work has been funded by the French National Research Agency (ANR) and by the German Research Foundation (DFG) joined project ANR-19-CE31-0026. GS is funded by a FFABR (Finanziamento delle Attivita` Base di Ricerca) grant of MIUR (Ministero dell’Istruzione, dell’Universita` e della Ricerca) and by a RFO research grant of DIFA (Diparti- mento di Fisica e Astronomia ‘Augusto Righi’) of the Alma Mater Studiorum Universita` di Bologna.

\bibliography{blibliography}{}
\bibliographystyle{aasjournal}
\newpage

\appendix
\renewcommand{\thefigure}{A\arabic{figure}}
\setcounter{figure}{0}
\renewcommand{\thetable}{A\arabic{table}}
\setcounter{table}{0}
\section{Incompressible model assumption}

\label{appendixA}
\begin{figure} 
    \centering
    \includegraphics{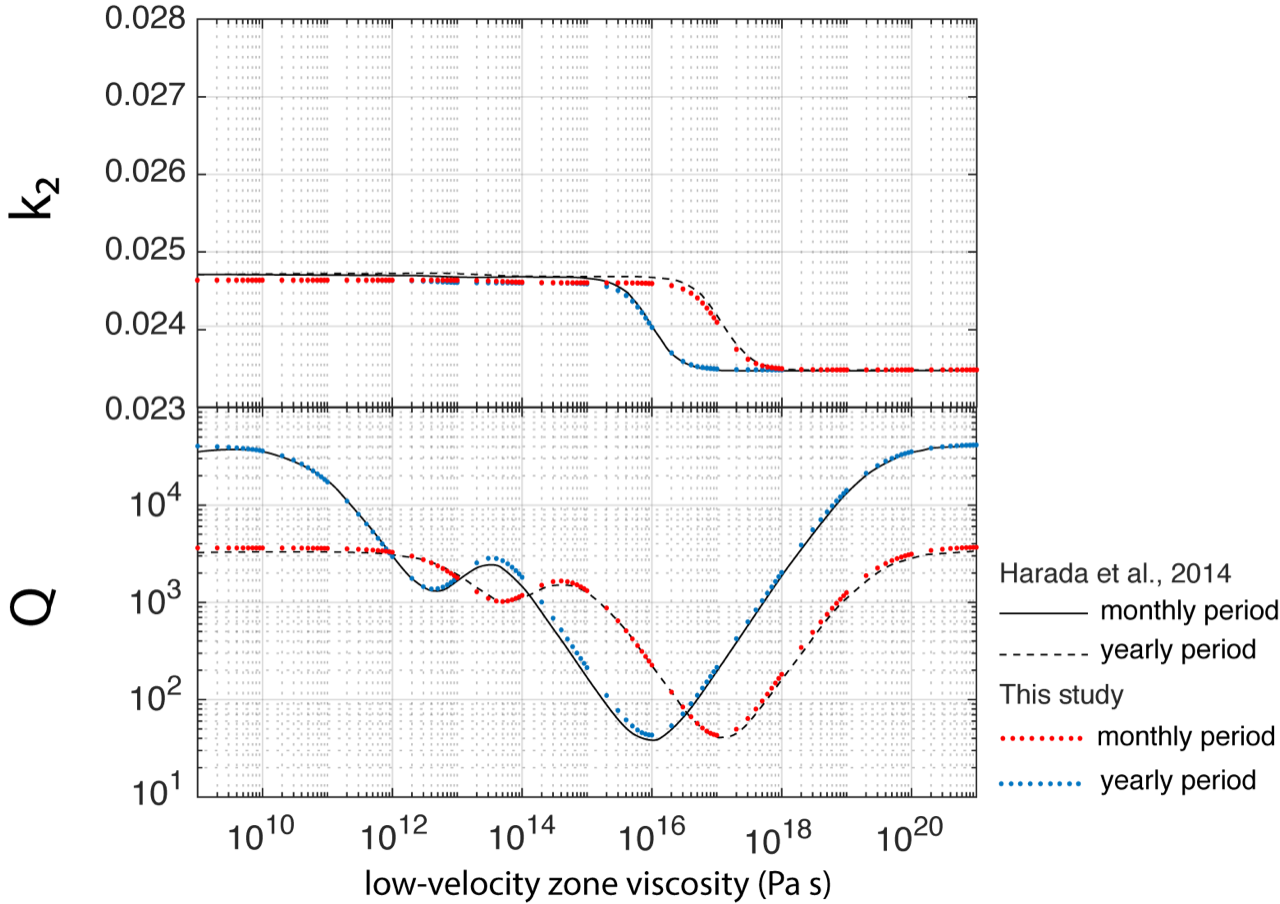}
    \caption{Comparisons between compressible and incompressible models for a Maxwell rheology. Top: $k_2$ as a function of the low viscosity zone (Pa$\cdot$s). Bottom: $Q$ as a function of the low viscosity zone (Pa$\cdot$s). Black lines correspond to the model of \citet{harada2014strong} while blue and red dots correspond to the $ALMA$ models for the two periods of interest.}
    \label{fig:A2}
\end{figure}

To estimate the impact of the incompressibility approximation assumed in this study, we have made some comparisons with the results provided by \citet{harada2014strong}, who employed a compressible Maxwell model. We took as inputs, an LVZ radius equal to 480 km and the same range of viscosity from 10$^9$ to 10$^{21}$ Pa$\cdot$s. For the density and rigidity, we used the values given by \citet{weber2011seismic} for each layer. Fig \ref{fig:A2} shows the dissipation $Q$ and the $k_{2}$ tidal Love Number as a function of the LVZ viscosity. The differences are negligible between the compressible model \citep{harada2014strong} and the incompressible (our work) as can also be seen from the numerical values listed in Table \ref{tab:TableA1}. Moreover, the differences are under the 3-sigma uncertainties given in Table 1 of this study. 

\begin{table}
\caption{Comparison between compressible model of \citep{harada2014strong} and the $ALMA$ incompressible model.}

    \centering
        \begin{tabular}{c c c c}
\hline
 Symbols & This study & \citet{harada2014strong} & difference \\
\hline
\hline
 $k_{2 F}$ & 0.02370 & 0.02369 & 1$\times$$10^{-5}$ \\
 $k_{2 l'}$ & 0.02468 & 0.02468 &  9$\times$$10^{-4}$\\
 $Q_F$ & 53 & 51 & 2 \\
 $Q_{l'}$ & 119 & 110 & 9\\
            \hline
        \end{tabular}
        \label{tab:TableA1}
\end{table}

\renewcommand{\thefigure}{B\arabic{figure}}
\setcounter{figure}{0}
\renewcommand{\thetable}{B\arabic{table}}
\setcounter{table}{0}

\section{K-means Algorithm}
\label{appendixB}
Here, we present the  $k$-means algorithm, which is an iterative, data-partitioning algorithm that assigns $N$ observations to one of $M$ clusters defined by centroids, where $k$ is chosen before the algorithm starts.
The data sets $X$ of our two model groups ($Category\,4$ and $Category\,5$) can be expressed as:

\begin{equation}
    X=\{x_1,...,x_N\}, x_N \in R^d,
\end{equation}

\begin{figure}
    \centering
    \includegraphics[scale=0.5]{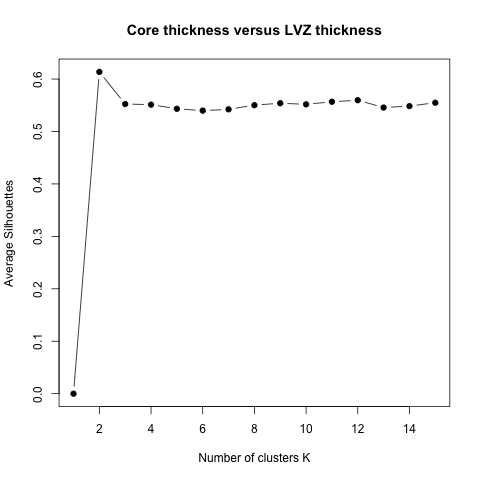}
    \caption{Silhouette coefficient as a function of the number of clusters for the outer core thicknesses versus LVZ thicknesses of $Category\,4$.}
    \label{fig:ann1}
\end{figure}

\begin{figure}
    \centering
    \includegraphics[scale=0.35]{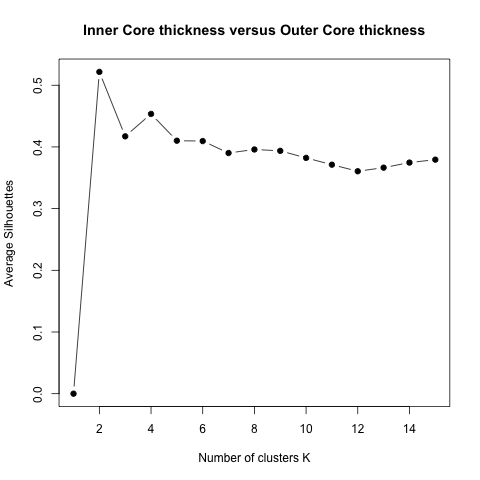}\includegraphics[scale=0.35]{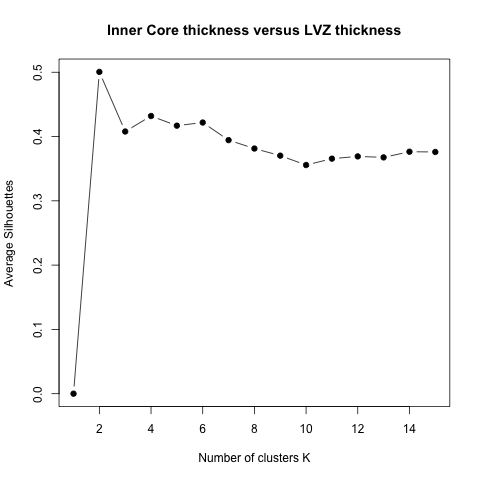}\includegraphics[scale=0.35]{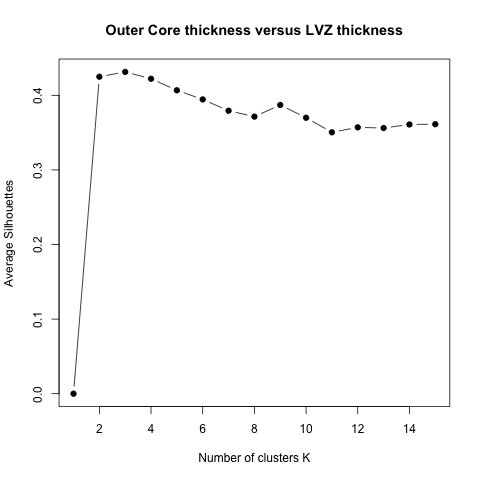}
    \caption{Evolution of the Silhouette coefficient relative to the number of clusters for the inner core versus outer core thicknesses (left-hand side), the Inner Core versus LVZ  thicknesses (middle), and the Outer Core versus LVZ  thicknesses (right-hand side) of $Category\,5$.}
    \label{fig:ann2}
\end{figure}

Where $x_n$ is a vector containing the (independent) parameters of each model, $N$ is the number of models and $d$ is the number of parameters for each model ($N=1126$ and $N=962$ for models with an inner core ($Category\,5$) and without an inner core ($Category\,4$), respectively). The $k$-means algorithm aims at partitioning the data set into $M$ disjoint sub-categories ($e.g.$, clusters) $C_1$,...,$C_M$, such that a clustering criterion is optimized. The commonly used criteria are the minimization of the sum of the squared Euclidian distances between each data point $x_i$ and the centroid $m_k$ of the sub-category $C_k$ (\citet{hartigan1979algorithm}):

\begin{equation}
    E(m_1,...,m_M) = \sum_{i=1}^{N} \sum_{k=1}^{M} I(F_i\in C_k) ||F_i-m_k||^2,
\end{equation}
Where $E$ is called \emph{clustering error} and $I(F)=1$ if $F$ is true and $0$ otherwise. 
\\
To determine the optimal number $M$ of clusters, we have proceeded clustering with $M$, the number of clusters varying from 1 to 10. We then compute the Silhouette parameter $S$ for each clustering and plot it against $M$. 
The Silhouette coefficient is defined by the following equation
\begin{equation}
S(i)=\frac{b(x_i)-a(x_i)}{max(a(x_i);b(x_i))}
\end{equation}
where $x_i$ is the member of one of the clusters, $a(x_i)$ is the average Euclidian distance between $x_i$ and all other members of the cluster to which $x_i$ belongs, and $b(x_i)$ is the average distance from $x_i$ to all clusters to which $x_i$ does not belong.
The optimum number of clusters is reached when the averaged $S$ reaches its maximum for a given number of clusters. 
One can see in Fig. \ref{fig:ann1} and Fig. \ref{fig:ann2} that for the five possible combinations ($Category\,4$ and $Category\,5$) $S$ reaches its maximum for $k=2$ except for the outer core thickness versus LVZ thickness. In this case, there is a plateau of a maximum between 2 and 3 possible clusters. However as for the others cases, the maximum is clearly reached at $k=2$, we keep results obtained with 2 clusterings.

\begin{figure}
    \centering
    \includegraphics[scale=0.45]{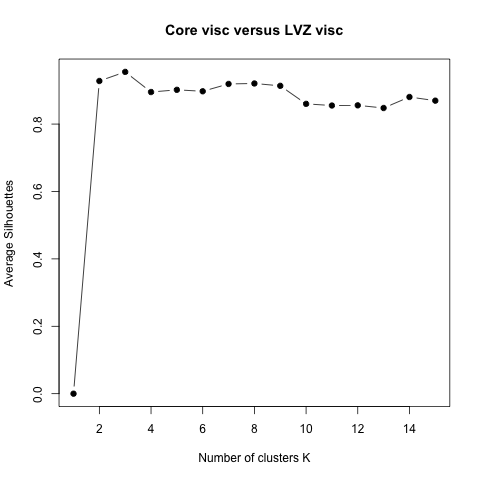}\includegraphics[scale=0.45]{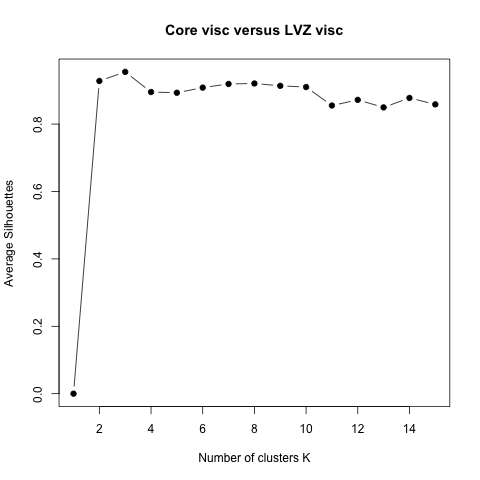}
    \caption{Evolution of the Silhouette coefficient relative to the number of clusters for the viscosity of the core versus LVZ $Category\,4$ (left-hand side) and of the outer core versus LVZ of $Category\,5$ (right-hand side).}
    \label{fig:ann3}
\end{figure}

In Fig. \ref{fig:ann3}, we plot the Silhouette average $S$ for the viscosity of the outer core and LVZ for $Category\,5$. It appears that the values stay very close to 1, indicating that the distribution is not clustered. Based on this result,  we do not consider clusters for the viscosities. Another indicator of clustering is the distribution of the 1-D histograms of Fig. \ref{fig:2Dhistoa}-b and \ref{fig:2Dhistob}-f. In these histograms, only one peak of density is visible together with a long tail. This type of distribution is also a good indication that there is {only one} density concentration for the viscosity distributions of both categories.

\renewcommand{\thefigure}{C\arabic{figure}}
\setcounter{figure}{0}
\renewcommand{\thetable}{C\arabic{table}}
\setcounter{table}{0}
\section{Other possible rheologies for the fluid core}
\label{appendixC}

\begin{figure} 
        \centering
        \includegraphics[scale=0.6]{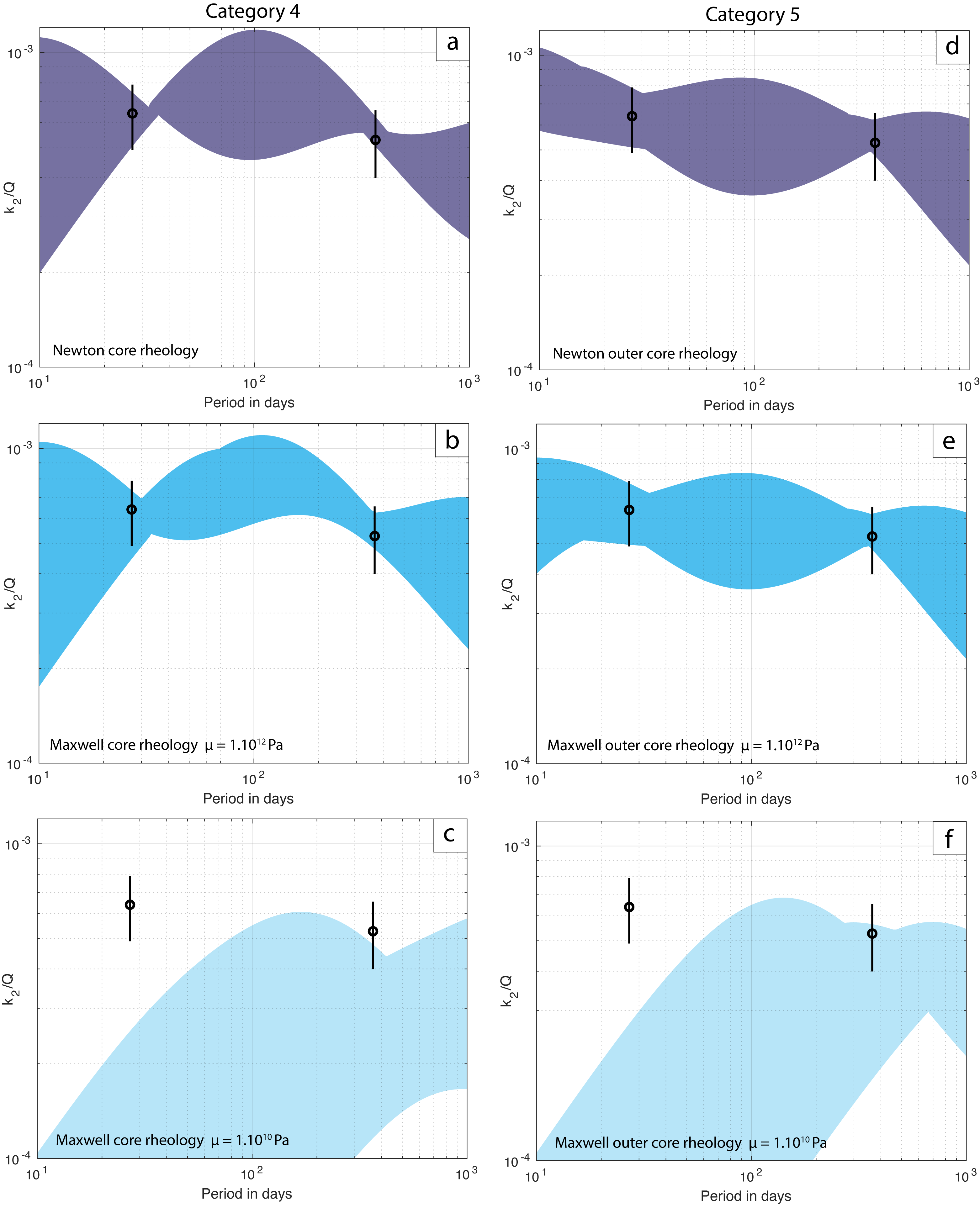}
        \caption{Examples of variations of the $k_2$/$Q$ parameter versus excitation periods in days obtained with $ALMA$$^3$ for randomly sampled profiles of $Category\,4$ (left-hand side) and $Category\,5$ (right-hand side) and different core rheologies: the Newtonian rheology in panels (a, d), the Maxwell rheology of  $\mu$= 10$^{10}$ Pa and $\mu$= 10$^{12}$ Pa in panels (b, e) and (c, f), respectively.}
\label{fig:Referee}
\end{figure}

We have tested the hypothesis of having a Maxwell instead of a Newton rheology in the core. To do so, we implemented ALMA$^3$, a core with a Maxwell rheology and we estimated TLNs and quality factors. As initial conditions, we have randomly sampled ten models (see Table. \ref{tab:tab_c1}) of $Category\,4$ and $Category\,5$. Since the rigidity of the fluid core is not constrained we thus used a range from low rigidity ($\mu$=1$\times10^{10}$ Pa) to unrealistic high rigidity ($\mu$=$1\times10^{12}$ Pa). The results are shown in Fig. \ref{fig:Referee}. In this Figure, one can see how the $k_2$/$Q$ changes with the rigidity as well as the observed $k_2$/$Q$ for the monthly and yearly period, here represented with dots and error bars. It is then visible that the Newtonian fluid core matches with the two observational constraints at the periods of interest while models with low rigidities do not fit with the observations. Only models with very high rigidities ($\mu$=$1\times 10^{12}$ Pa) may fit with the observations. For these cases, the required rigidities are higher than the one used for the inner core and low-velocity zone of $4.23\times10^{10}$ Pa and $2.48\times10^{10}$ Pa, respectively.

\begin{table}[ht]
        \centering

\caption{Characteristics of the randomly selected models. Upper and lower scripts refer to maximum minimum values, respectively.}
            \begin{tabular}{c c c c}
            Symbol & Unit & Category 4 & Category 5 \\
            \hline
            \hline
            {$R_{LVZ}$} & km & $^{501}_{497}$ & $^{501}_{498}$ \\
            {$R_{C}$} & km & $^{413}_{357}$ & -- \\
            {$R_{OC}$} &  km & -- & $^{423}_{327}$\\
            {$R_{IC}$} & km & -- &  $^{243}_{191}$ \\
            {$\eta_{LVZ}$} & Pa$\cdot$s & $^{17.00}_{15.77}$ & $^{18.00}_{16.47}$\\
            {$\eta_{C}$} & Pa$\cdot$s  & $^{25.84}_{16.69}$& -- \\
            {$\eta_{OC}$} & Pa$\cdot$s  & -- & $^{17.00}_{15.84}$ \\
             \hline
\end{tabular}
\label{tab:tab_c1}
\end{table}
\end{document}